\documentclass{article}
\usepackage{arxiv}
\usepackage{longtable}                        
\usepackage[utf8]{inputenc}
\usepackage[T1]{fontenc}   
\usepackage{hyperref}      
\usepackage{url}           
\usepackage{booktabs}       
\usepackage{amsfonts}       
\usepackage{nicefrac}       
\usepackage{bm}
\usepackage{natbib}
\usepackage{lipsum}
\usepackage{graphicx}
\usepackage{amsmath}
\usepackage{amssymb}
\usepackage{algorithm}
\usepackage{algpseudocode}
\usepackage{adjustbox}
\usepackage{mathtools}
\usepackage{fullpage}
\usepackage[dvipsnames]{xcolor}
\usepackage{comment}
\usepackage{tikz}
\usepackage{pgffor}
\usetikzlibrary{bayesnet}

\usepackage{setspace}
\newcommand{\ALB}{AA}
\newcommand{\NYU}{AP}
\newcommand{\UAB}{AV}
\newcommand{\UCSF}{AW}

\newcommand{\schwarz}{\color{black}}
\newcommand{\orange}{\color{orange}}

\newcommand{\bchGP}{\orange \em}

\newcommand{\ech}{\schwarz \rm}

\renewcommand{\th}{\theta}
\newcommand{\sig}{\sigma}
\newcommand{\deltah}{\bm{\widehat{\delta}}_{EB}}

\newcommand{\thb}{\bm{\theta}}
\newcommand{\rhob}{\bm{\rho}}
\newcommand{\rhobm}{\bm{\rho}_{\text{-}i}}

\newcommand{\deltab}{\bm{\delta}}

\newcommand{\tb}{\bm{\theta}}
\newcommand{\xib}{\bm{\xi}}

\newcommand{\xb}{\bm{x}}
\newcommand{\yb}{\bm{y}}
\newcommand{\zb}{\bm{z}}

\newcommand{\Rb}{R}

\newcommand{\zerob}{\bm{0}}

\newcommand{\tbt}{\bm{\widetilde{\tb}}}
\newcommand{\Rbt}{\widetilde{R}}

\newcommand{\sm}{^{\star-}}
\newcommand{\mm}{^{-}}

\newcommand{\xibs}{\bm{\xi^\star}}
\newcommand{\xis}{\xi^\star}

\newcommand{\vepsb}{\bm{\varepsilon}}

\newcommand{\TT}{\mathcal{T}}
\newcommand{\EE}{\mathbb{E}}
\newcommand{\II}{\mathbb{I}}
\newcommand{\AAA}{\mathcal{A}}
\newcommand{\Pxig}{P_{\xi,g}}
\newcommand{\g}{g_i}
\newcommand{\gl}{g_{\ell}^{(i)}}
\newcommand{\glp}{g_{\ell'}^{(i)}}
\newcommand{\pl}{p_{\ell}^{(i)}}
\newcommand{\plp}{p_{\ell'}^{(i)}}
\newcommand{\Gm}{G_{i,t}}
\newcommand{\Wm}{W_{i,t}}

\newcommand{\rell}{\ell}
\newcommand{\rellp}{\ell'}

\newcommand{\zerop}{\phantom{0}}

\newcommand{\Ber}{\text{Ber}}
\newcommand{\N}{\mathcal{N}}
\newcommand{\DD}{\mathcal{D}}
\newcommand{\SMC}{\text{SMC}}

\newcommand{\PY}{\text{PY}}
\newcommand{\HEP}{\text{HEP}}
\newcommand{\SUN}{\text{SUN}}
\newcommand{\MCMC}{\text{MCMC}}
\newcommand{\Dir}{\text{Dir}}

\newcommand{\rhom}{\bm{\rho_{\text{-}i}}}
\newcommand{\xibm}{\bm{\xi}_{\text{-}i}}
\newcommand{\rhomn}{\bm{\rho_{\text{-}n}}}
\newcommand{\xibmn}{\bm{\xi}_{\text{-}n}}

\newcommand{\rhot}{\widetilde{\rho}}

\title{Borrowing strength between unaligned binary time-series via Bayesian nonparametric rescaling of Unified Skewed Normal priors}

\author{
Beatrice Cantoni\\
Department of Statistics and Data Science, University of Texas at Austin, Austin, Texas, U.S.A.\\
\texttt{beatrice.cantoni@utexas.edu}
\AND
Giovanni Poli\\
Department of Statistics, Computer Science,
Applications “G. Parenti”,
University of Florence, Florence, Italy.\\
\texttt{giovanni.poli@unifi.it }
\AND
Elizabeth Juarez-Colunga\\
Department of Biostatistics,\\
University of Colorado, Denver, USA.\\
\AND
Peter M\"uller \\
Department of Statistics and Data Science, University of Texas at Austin, Austin, Texas, U.S.A.\\
\texttt{pmueller@math.utexas.edu}}

\begin{document}
% \thanks{B.C. and G.P. are equal first authors}

\maketitle

\begin{abstract}
  We define a Bayesian semi-parametric model to effectively conduct inference with unaligned longitudinal binary data. The proposed strategy is motivated by data from the Human Epilepsy Project (HEP), which collects seizure occurrence data for epilepsy patients, together with relevant covariates. The model is designed to flexibly accommodate the particular challenges that arise with such data. First, epilepsy data require models that can allow for extensive heterogeneity, across both patients and time. With this regard,  state space models offer a flexible, yet still analytically amenable  class of models. 
  Nevertheless, seizure time-series might share similar behavioral patterns, such as local prolonged periods of elevated seizure presence, which we refer to as \textit{clumping}. 
  Such similarities can  be used to share strength across patients and define subgroups. However, due to the lack of alignment, straightforward hierarchical modeling of latent state space parameters is not practicable. To overcome this constraint, we construct a strategy that preserves the flexibility of individual trajectories while also   exploiting similarities across individuals to borrow information through a nonparametric prior. On the one hand, heterogeneity is ensured by (\textit{almost}) subject-specific state-space submodels. On the other, borrowing of information is obtained by introducing a Pitman-Yor prior on group-specific probabilities for patterns of clinical interest.  We design a posterior sampling strategy that leverages recent  developments of binary state space models using the Unified Skewed Normal family (SUN). The model, which allows the sharing of information across individuals with similar disease traits over time, can more generally be adapted to any setting characterized by and unaligned binary longitudinal data.
\end{abstract}


\section{Introduction}\label{sec:intro}
We introduce a semi-parametric Bayesian model for inference with un-aligned longitudinal binary data. 
The model is designed for the recognition of patterns of clinical interest.
The main feature of the proposed approach is that it allows the inference on defined patterns to be informed, combining evidence from all patients while still providing the required marginal flexibility. 
This is done through a hierarchical model that enables the borrowing of strength using a flexible structure that can be tailored to specific inference goals and applications.
Inference exploits recently introduced computation-efficient posterior simulation algorithms for state-space models with binary outcomes, based on uniform skewed normal ($\SUN$) prior \citep{arellano2006unification}.

The proposed approach is motivated by data analysis for an epilepsy study. 
Epilepsy is recognized as one of the most common neurological diseases, affecting around 3 million individuals in the United States and approximately 50 million people worldwide (\citealp{world2019epilepsy,WebHEP}).
Studies aimed at better understanding and controlling this condition are therefore of crucial importance. Despite growing interest in focal epilepsy and extensive related research, there are only a few prospective studies capturing detailed information on the timing and recurrence of seizures, as in the case of the Human Epilepsy
Project ($\HEP$). Our work was indeed motivated by data from $\HEP$, an ongoing 3-year study that collects prospective follow-up for 450 newly treated patients, recording seizure episodes,
constant subject characteristics, and time-varying covariates. A goal of $\HEP$ is to collect data that can support the development of models that meaningfully  profile new patients, laying the foundation for forecasting disease progression and informing optimal treatment recommendations.

Suitable models for epilepsy time-series should allow for a high level of flexibility in capturing marginal individual trajectories over time. This flexibility is well captured by a state space model with an adequate auto-regressive component. Such a model also allows for the inclusion of time-varying covariates, particularly treatment indicators. Furthermore, epilepsy is known to affect patients by a set of behavioral patterns that might be shared among groups of individuals. This motivates interest in identifying population subgroups and in using model features that facilitate the borrowing of information across individuals within the same subgroup. However, the lack of time alignment across individuals, even if they present similar pattern features that occur at different times in their observation period, combined with high levels of randomness within and between patients, hinders the use of hierarchical models that rely on matching individuals by pairing their parameters. Currently, available inference models for epilepsy data do not fully address these challenges. Although epilepsy as a dynamic disease has been discussed in the literature, existing strategies generally focus on aggregate events (e.g. \cite{chiang2018dyepilepsy} and \cite{KristenMiller}), or the use hidden Markov models with discrete latent states (e.g., \citealt{wang2023bayesian}).
To our knowledge, only one existing approach models continuous latent states for epilepsy while accommodating the heterogeneity of subject-specific trajectories \citep{wang2022bayesian}. That approach implements inference at the population level across subjects as post-processing of marginals posterior distributions. In contrast, our approach seeks to incorporate population-level inference directly within a hierarchical model, allowing for borrowing of information across individuals. Relatedly, dynamic prediction models have been used successfully in other clinical areas besides epilepsy \citep{van2008dynamic, rizopoulos2011dynamic,
  taylor2013real, mauguen2013dynamic, garcia2021dynamic}.
However, such models typically do not support the identification of subgroups of trajectories, as we hypothesize exist within the population patients with epilepsy.

We propose an approach based on appropriately parameterized Bayesian dynamic state-space models \citep{prado2010time}.
The outcomes are modeled using a probit sampling model.
Recent work has introduced computationally efficient ways to extend this class of models to binary outcomes \citep{durante2019conjugate,fasano2021closed,anceschi2023bayesian}, allowing the modeling of such time-series through an underlying sequence of continuous states.  
Such models, when properly parameterized, provide the flexibility needed for modeling epilepsy seizures, and define reliable priors forsubject- and time- specific parameters that represent the dynamic latent state for the daily risk of seizure. 
In our application, the linear predictor on the probit-scale includes a regression with dynamic time-varying coefficients for time-varying covariates and static coefficients for subject-specific baseline covariates. 
Posterior inference under this model specification can be implemented analytically as the $\SUN$ prior on the latent states and the probit likelihood are conjugate \citep{durante2019conjugate}.
However, closed-form inference is not practically feasible due to an explosion of latent dimensions.
Instead \citep{fasano2021closed,anceschi2023bayesian} introduced a sequential Monte Carlo ($\SMC$) method for efficient posterior simulation for time-varying parameters.
We further develop and extend this $\SMC$ scheme to accommodate static parameters in the transition distribution of the state space model, and to allow for missing responses. 
If on the one hand \ech the flexibility of the state space model allows to effectively fit the data for each patient, on the other it complicates the definition and reporting of inference summaries for the study population and the sharing of information across subjects. 
To mitigate this complication, we integrate pre-defined patterns of clinical interest in the probabilistic modeling and assume that clusters of subjects have different probabilities of experiencing such patterns in their longitudinal trajectories. 
In particular, such patterns of clinical interest are defined by suitable summaries of each subject's latent dynamic state variables, and clusters are defined by shared probabilities for such patterns. 
Note that the patterns themselves are not exclusive to specific clusters.
Clusters are identified on the basis of probabilities for patterns.
We use a Pitman-Yor ($\PY$) process \citep{pitman1997two} to define a prior for the cluster arrangements.
Note that although the described model is motivated by the application to the $\HEP$ data, similar features are common to other instances of binary panel data, making the proposed approach suitable for a variety of applications. 

In Section \ref{sec:data}, we describe in detail the 
motivating application and our data.
In Section \ref{sec:model}, we describe the hierarchical structure and logic behind the modeling choices. 
Section \ref{sec:priorelicitation} discusses prior elicitation and modeling choices for the application.   
In Section \ref{sec:inference}, we describe posterior simulation strategies for the described model. 
In Section \ref{sec:sim}, we assess the proposed inference approach with a small simulation study. 
In Section \ref{sec:Application} we show results for the $\HEP$ data, and Section \ref{sec:conclusion} concludes with some final comments.

\section{The Human Epilepsy Project Data}\label{sec:data}

This section outlines the data obtained through Human Epilepsy Project ($\HEP$) to frame the project within medical research and to highlight the methodological challenges that motivate this manuscript.
The first cohort of patients was a prospective 3-year follow-up study of $450$ newly treated patients of various ages, which terminated in 2020; for simplicity we will refer to this cohort simply as HEP. 
Patients were asked to daily self-report seizure event occurrence. As a result, the dataset consists of binary time series for each patient, indicating the presence or absence of at least one seizure per each day (or \texttt{NA} for missing responses). 
Data collection included demographic and clinical covariates at baseline, and information on antiepileptic drug treatments over time.   
In this study, we exclude the few individuals reporting data for less than 30 days, and we analyze data from $N=390$ patients. 
The study data consisted of 238 (61.03\%) females, had a median age of 31 (20-42).
%Additional demographics and baseline characteristics are reported in XXX.
Most individuals were started on Leviteracetam at enrollment (39 \%), though treatment regimes varied over time. 
In total, 33 different combinations of antiepileptic medications were recorded and included in the analysis. The time series range from 34 to 1094 days in length, with an average duration of 733.07 days, making it impractical to align them or derive common interpretations based on time indices.

An important long term goal of the HEP study is to predict seizure trajectories, and identify  meaningful patterns over time. Intermediate goals include the development of automated systems to identify unknown subgroups, enabling the borrowing of information across similar patients. Of great clinical interest is also the early identification of treatment-resistant patients, those unlikely to achieve seizure remission through continued medication trials \citep{schiller2008quantifying}.  The true magnitude of treatment-resistant epilepsy,
particularly in specific forms of epilepsy,  remains unclear because of methodological limitations of previous studies, such as failure to account for variability in baseline seizure frequency.   There is particular interest in understanding whether clinical characteristics of seizure occurrence, such as seizure pattern, time to next seizure after treatment, or seizure clustering can help predict treatment resistance.

$\HEP$ data are broadly described as binary longitudinal time series. Several key features must be considered in model development. First, data exhibit substantial heterogeneity, both within each patient over time and across different patients. For example, seizure frequency varies widely between individuals. It can also change over time for the same individual.% (see the dots in Figure \ref{fig:marginal} for examples of data trajectories of 3 individuals). 
This variability creates a need for flexible models that accommodate both between and within differences, especially given the long follow-up time for some patients. Another important feature of the $\HEP$ data is the time-varying nature of antiseizure medication covariates. A suitable model needs to allow for static as well as time-varying covariates ($\xb_i$ and $\zb_{it}$, respectively, in the upcoming modeling section).  Finally, epilepsy is known to manifest in distinct temporal patterns across patients. The occurrence or propensity of such patterns across different patients can be used to identify homogeneous subgroups of patients in the population \citep{KristenMiller}. Examples of patterns include (i) what we shall refer to as \textit{clumping}, a self-exciting process feature where a  seizure episode increases the likelihood of additional seizures shortly afterward (e.g. a triggering event), or (ii) a specific behavior for the time-series within a specific time window. These features motivate the need for statistical inference methods to discover and characterize such subpopulations and borrow strength within such subpopulations. The proposed modeling approach, discussed in the next subsection, focuses on identifying clinically relevant patterns that characterize the time series both globally and locally.

\section{Model-based Clustering of Unaligned Time-Series}\label{sec:model}
\subsection{Hierarchical Prior and the Scaled Model}
\label{subsec:hiermod}
We introduce some notation.
The data are binary time series $\DD=\{\yb_i:\ i=1,\ldots,n\}$ for patient $i$ with $\yb_i =\{y_{i,t}:\ t=1,\ldots,T_i\}$.  
Here $y_{i,t}\in\{0,1\}$ is a binary indicator for the occurrence of
a seizure on day $t$.
We assume a probit sampling model $y_{i,t} \mid \deltab,\theta_{i,t}
\sim \Ber\big(\Phi^{-1}(\xb_i^\top\deltab+\zb_{i,t}^\top\theta_{i,t})\big)$.
The latent probit score is decomposed into two terms.
The first term, $\mu_i=\xb_i^\top\deltab$, defines a regression on a
vector of time-constant  baseline covariates $\xb_i$. 
The second term
introduces a subject- and time-specific random effect as a regression
on time-varying covariates (or design vectors) $\zb_{i,t}$ as
$\gamma_{i,t}=\zb_{i,t}^\top\thb_{i,t}$.
The combination of time-varying and static parameters, as
well as the use of subject-specific and population
parameters models introduce the desired 
flexibility and heterogeneity over time and patients. 
To ensure the intended borrowing of information, we introduce a
hierarchical prior on parameters $\tb_i=\{\tb_{i,t}:\ t=1,\ldots,
T_i\}$, which characterizes the time-varying random effects.   
One inference aim is to recognize clusters of patients with similar behavioral patterns. 
A complication arises from the lack of time alignment across patients.

In anticipation of the construction used to overcome this challenge
and to allow for a clustering strategy based on similarities of the
implied trajectories, we introduce pre-defined events of clinical interest.
Assume there are $L$ events or combinations of events of
interest. We introduce a discrete random variable
$\Rb_i \in \{0,\ldots,L-1\}$ to record the events for patient
$i$.
The events are defined in terms of the subject's marginal time series
$\tb_i$,  typically by way of some relevant summary statistic
$S(\tb_i)$, such as a threshold on the average fitted probability of
$y_{it}=1$. We will discuss specific examples later. 
Next, we introduce $\xib_i$ as an $L$-dimensional vector of
probabilities $\xi_{i,\ell}=\Pr(\Rb_i=\rell)$. 
The strategy is to group patients according to $\xib_i$,  by
assuming $\xib_i \sim P$ for a discrete random probability measure
$P$.
Ties among the $\xib_i$ define the desired clusters.
Let $\{\xibs_1,\ldots,\xibs_K\}$ denote the unique values among the
$\xib_i$. Patient $i$ is in cluster $k$ if $\xib_i=\xibs_k$.
Similar constructions for random partitions based on 
sampling from a discrete probability measure are commonly used in
Bayesian inference, due to
a representation theorem by \cite{kingman1978representation}. 

\begin{figure}[t]
    \centering    \includegraphics[width = \textwidth]{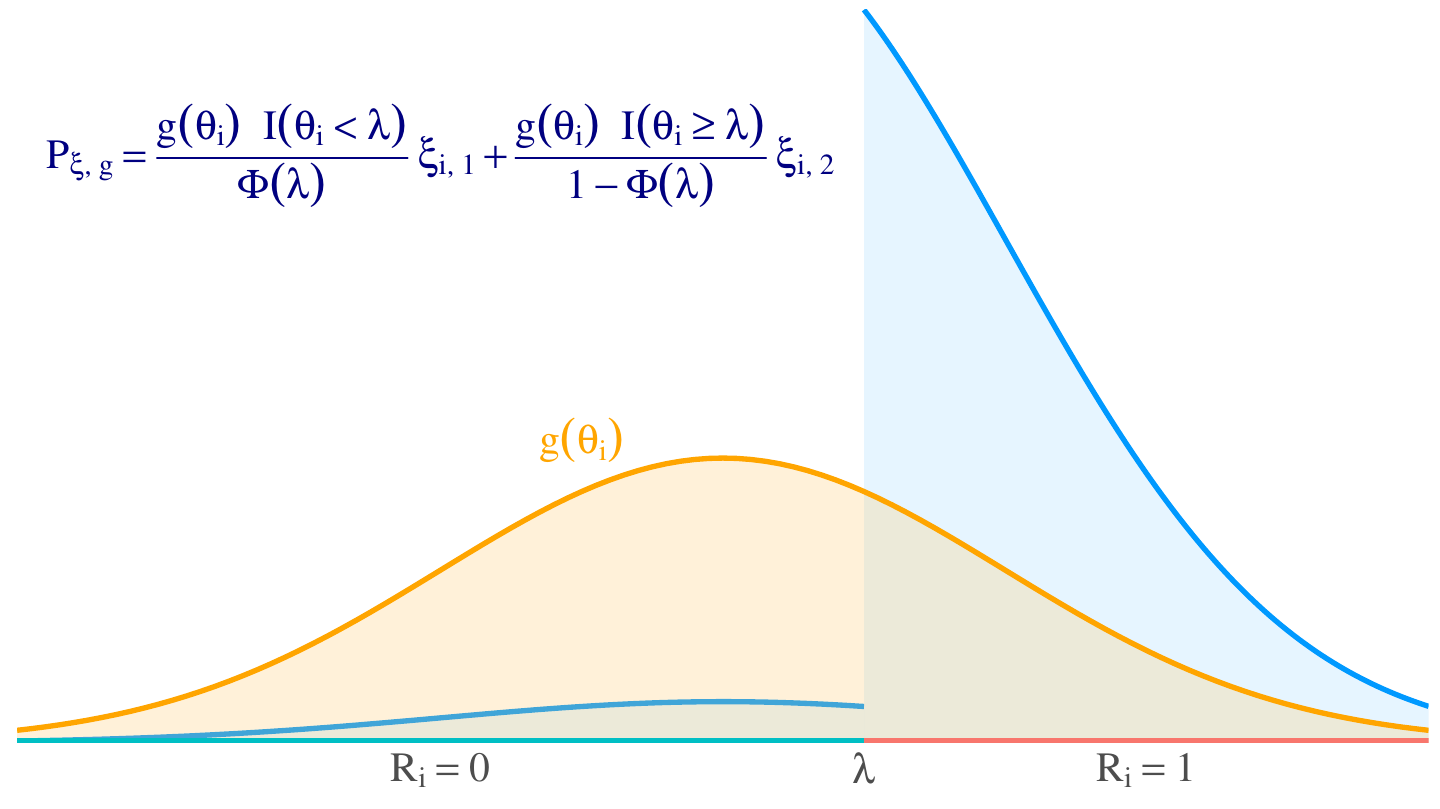}
    \caption{Example figure of the scaled model.
    The example compares the scaled distribution with the reference distribution.
    It assumes $g(\theta)=N(\theta\mid\mu,\sigma^2)$, $\xib_1=(0.1,0.9)^\top$ and $\Rb_1=\II(\theta\geq\lambda)$.}
    \label{fig:eExample}
\end{figure}

We propose
\begin{align}\label{eq:thxi}
  p(\tb_i,\xib_i\mid P)
  &= p(\tb_i\mid\Rb_i=\rell)\
    p(\Rb_i= \rell \mid  \xib_i)\  p(\xib_i\mid P),\\
  & = p(\tb_i\mid\Rb_i=\rell)\, \xi_{i,\ell}\, P(\xib_i) \nonumber
\end{align}
Let $\Theta_i$ denote the support of $\tb_i$, and let
$\AAA_\ell = \{\tb_i\in\Theta_i:\; R_i=\ell \}$
(for simplicity suppressing an additional index $i$ in $\AAA_\ell$ that subject specific, as the sets $\Theta_i$ vary with $T_i$). 
The first factor in the proposed inference model \eqref{eq:thxi} is
constructed using $\Rb_i$ and a reference model $\g(\tb_i)$ (to be introduced later) as
\begin{equation}\label{scaledvsunscaled}
  p(\tb_i\mid \Rb_i=\rell) \equiv \g(\tb_i\mid
  \Rb_i=\rell)=\frac{\g(\tb_i)}{\g(\Rb_i=\rell)}\cdot\II(\tb_i\in \AAA_\ell).
\end{equation}
In other words, the statistic $S(\thb)$ maps the subject-specific parameters ($\thb_i$) into subsets of their sample space ($\mathcal{A_\ell}$). The latter are characterized by a clinical interpretation valid throughout the population \ech (e.g. ascending or descending trends for the time series). 
The implied prior $p(\tb_i \mid \xib_i)$,is then
$$
\Pxig \equiv p(\tb_i \mid \xib_i)=\sum_{\ell=0}^{L-1}\ p(\tb_i \mid
\Rb_i=\rell)\ p(\Rb_i=\ell\mid \xib_i) = \sum_{\ell=0}^{L-1}\  \g(\tb_i \mid
\Rb_i=\rell)\cdot\xi_{i\ell}
$$
with cluster-specific parameters $\xi_i$.
We refer to $\Pxig$ as the \textit{scaled} model, i.e., a rescaled
version of $\g$ to ensure, by construction,  
$p(\Rb_i=\rell\mid \xib_i)=\xi_{i,\ell}$. 
In Section \ref{sec:priorelicitation}, we will specify and motivate
our choice for $\g$.  
Figure \ref{fig:eExample} shows a stylized (univariate)
illustration of $\g(\th_i)$ and $\Pxig$. In this cartoon we
assume $g_1(\theta_1)=N(\theta_1\mid\mu,\sigma^2)$ and
$\xib_1=(0.1,0.9)^\top$,
$\Rb_i=\II(\theta_i\geq\lambda)$ and
compare the \textit{scaled} distribution $\Pxig$ with the (unscaled)
reference distribution $\g(\th)$.
The illustration assumes one event of interest 
with $\Rb_i \in \{0,1\}$. 
The figure highlights how the construction induces
specific priors for $\thb_i$ by way of $\Rb_i$ which
defines cluster-specific probabilities for patterns of interest.

\begin{figure}[t]
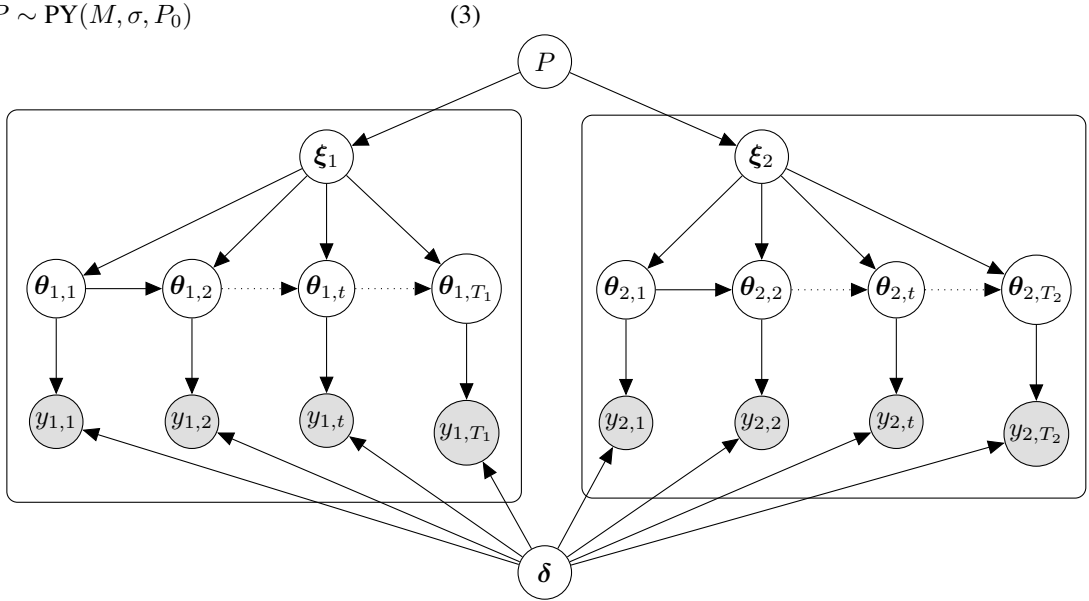

  \begin{minipage}[t]{.45\textwidth}
    \vspace{0pt}
  \begin{align}
    \intertext{Sampling model:}
    &y_{i,t} \mid  \deltab,  \theta_{i,t} \sim
      \Ber \left(\Phi \left(\xb_i^\top \deltab  +
      \zb_{i,t}^\top\theta_{i,t}\right)   \right)       \nonumber \\
    \intertext{Regression on baseline covariates:}
    &\deltab \sim N(m_{\delta}, V_{\delta}) \nonumber   \\
    \intertext{Dynamic state-space model -- system equation:}
    & \tb_i\mid \boldsymbol{\xi}_i \sim P_{\xi,g} \nonumber \\
    \intertext{Clustering:}
    &\xib_i=(\xi_{i,1},\dots,\xi_{i,L})^\top\mid P \sim P \nonumber \\
    \intertext{BNP hyperprior:}
    &P\sim \PY(M,\sig,P_0) \label{eq:py} % \label{eq:py}%\mathcal{SS}\big(\vartheta\big) 
\end{align}
\end{minipage}\hspace{.2cm}
  \begin{minipage}[t]{.5\textwidth}
    \vspace{0pt}
\begin{align}
  \intertext{Patients:}
  &i=1,\ldots,n  \label{eq:p}\\
  \intertext{Times:}
  &t=1,\dots,T_i \nonumber \\
  \intertext{Cluster-specific state-space model:}
&P_{\xi,g}=\sum_{\ell=1}^L \dfrac{g(\tb_i)\cdot \II(\theta_i\in
                                                    \mathcal{A}_\ell)}{g(\Rb_i=\rell)}\cdot\xi_{i,\ell}\nonumber
\end{align}
\end{minipage}
\medskip
\begin{minipage}{\textwidth}
\begin{center}
\resizebox{.875\textwidth}{!}{
     \tikz{ %
\node[latent] (xi) {$\xib_1$} ; %
\node[latent, right= of xi, xshift = 4 cm] (xip) {$\xib_{2}$} ; %
\node[latent, below=of xi] (theta) {$\tb_{1,t}$} ; %
\node[latent, left=of theta] (theta2) {$\tb_{1,2}$} ; %
\node[latent, left=of theta2] (theta1) {$\tb_{1,1}$} ; %
\node[latent, right=of theta] (thetan) {$\tb_{1,T_1}$} ; %
\node[obs, below=of theta] (Y) {$y_{1,t}$} ; %
\node[obs, below=of theta1] (Y1) {$y_{1,1}$} ; %
\node[obs, below=of theta2] (Y2) {$y_{1,2}$} ; %
\node[obs, below=of thetan] (Yn) {$y_{1,T_1}$} ; %
\node[latent, below=of xip] (theta2p) {$\tb_{2,2}$} ; %
\node[latent, left =of theta2p] (theta1p) {$\tb_{2,1}$} ; %
\node[latent, right =of theta2p] (thetatp) {$\tb_{2,t}$} ; %
\node[latent, right =of thetatp] (thetaTp) {$\tb_{2,T_{2}}$} ; %
\node[latent, right= of xi, xshift = 1.15 cm, yshift = 1.25 cm] (P) {$P$} ; %
\node[latent, below = of P, yshift = -5.0 cm] (delta) {$\deltab$} ; %
\node[obs, below=of thetatp] (Yp) {$y_{2,t}$} ; %
\node[obs, below=of theta1p] (Y1p) {$y_{2,1}$} ; %
\node[obs, below=of theta2p] (Y2p) {$y_{2,2}$} ; %
\node[obs, below=of thetaTp] (Ynp) {$y_{2,T_2}$} ; %
\edge {xi} {theta,theta1,theta2,thetan} ; %
\edge {theta1}{Y1} ; %
\edge {theta2} {Y2} ; %
\edge {theta1} {theta2} ; %
\edge[dotted] {theta2} {theta} ; %
\edge[dotted] {theta} {thetan} ; %
\edge {theta} {Y} ; %
\edge {thetan} {Yn} ; %
\edge {delta} {Y1,Y2,Y,Yn,Y1p,Y2p,Yp,Ynp} ; %
\edge {P} {xi,xip} ; %
\edge {xip} {theta1p,thetatp,theta2p,thetaTp} ; %
\edge {theta1p}{Y1p} ; %
\edge {theta2p} {Y2p} ; %
\edge {thetatp}{Yp} ; %
\edge {thetaTp} {Ynp} ; %
\edge {theta1p} {theta2p} ; %
\edge[dotted] {theta2p} {thetatp} ; %
\edge[dotted] {thetatp} {thetaTp} ; %
\plate[inner sep=0.25cm] {plate1} {(Y1)(theta1)(Y2)(theta2)(Y)(theta)(Yn)(thetan)(xi)} {$  $ }; %
\plate[inner sep=0.18cm] {plate2} {(Y1p)(theta1p)(Y2p)(theta2p)(Yp)(thetatp)(Ynp)(thetaTp)(xip)} {$  $ }; %
  }}
\end{center}
\end{minipage}
\caption{Directed Acyclic Graph representation for two subjects.}
\label{fig:DAG}
\end{figure}

The model is completed with a hyperprior on $P$,
which defines a random discrete probability measure.  
Priors on random probability measures are known in the literature
as nonparametric Bayesian inference ($\mbox{BNP}$)
\citep{ghosal2017fundamentals}. 
In particular, we use the  Pitman-Yor prior \citep{pitman1997two}.
See Section \ref{HEP:4.2} for motivations about the specific
elicitation of the priors $\g$ and $P$. 
Figure \ref{fig:DAG} summarizes the resulting hierarchical structure,
using as example the case of two subjects only. 
For reference, we state the complete inference model
(see the upcoming discussion for an explanation of the label ``dynamic
state space model'' for the prior on $\tb_i$).

\subsection{Remarks on the Hierarchical Structure}

We elaborate more on the intention behind the proposed hierarchical structure by considering two limiting cases. 
The first case assumes all singleton clusters, that is, the
number of clusters being equal to the number of subjects.
The second case assumes only one cluster.
Under the first scenario, the model reduces to subject-specific
scaling of the reference model $\g$. No information about the
probabilities $\xib_i$ of the different patterns is shared.
Under the second case instead, one common set of probabilities $\xib$ is
learned.   
As a consequence, rare behaviors are shrunk toward more common behaviors.
In contrast, our construction defines a compromise between
these two limiting behaviors by creating new clusters as
needed. 
Note that to be informative and interpretable, scaling
probabilities $\xib_i$ should be closer to the corners of the
simplex. 
We chose hyperpriors to favor such features by using Dirichlet
distributions with small hyperparameters. 

Some observations on the comparison of
the reference model $\g$ versus the marginal prior $\Pxig$  under the
inference model. 
First, note that the \textit{unscaled} reference
model $\g$ is elicited separately for each subject.
This allows, for example, a different number of treatments and
different length of follow-up for each patient.
Related to this observation, the denominator $\g(\Rb_i=\rell)$ of 
\eqref{scaledvsunscaled} is subject-specific.
The marginal model $\Pxig$ has a cluster-specific component ($\xib$)
and a subject-specific one ($\g$).
To highlight the difference between the reference model and the corresponding quantities under the inference model, we now focus on the
predictive distribution for $R_i$, for
subject $i=n$ given subjects $1,\ldots,n-1$. 
Note that under the reference model $\g$, there is no notion of
dependence between subjects.  
Thus, probabilities for the event $\Rb_n=\rell$ are
independent across subjects.
In contrast, the predictive distributions under the proposed 
inference model depends on the values
$\xib_1,\ldots,\xib_{n-1}$.
As we shall still discuss in more detail later (Section
\ref{sub:hier}), the PY prior
 with base measure $P_0$
in \eqref{eq:py} implies a discrete
probability measure $P$, which in turn implies positive probability
for ties among the $\xib_i$. 
Let then $\xibs_1, \ldots, \xibs_H$
denote the unique values among  $ \xibmn= \{\xib_1,\ldots,\xib_{n-1}\}$,
define cluster membership indicators 
$\rho_i=h$ if  $\xib_i=\xibs_h$, and 
let $\rhomn=(\rho_1, \ldots, \rho_{n\text{-}1})^\top$.
Then,  
\begin{equation}
\begin{split}
  p(\Rb_n=\rell\mid\xibmn,\rhomn)
  &=\sum_{h =1}^{H+1}p(\Rb_n=\rell\mid\rho_n=h, \xibmn)\
    p(\rho_n=h\mid\rhomn)= \\ 
  &= \sum_{h =1}^{H}\xi_{\ell,h}^\star\cdot p(\rho_n=h\mid\rhomn) +
    p(\rho_n=H+1\mid\rhomn)\cdot\int\xi_{\ell,i}\ dP_0(\xib_i),
\end{split}\label{pred}
\end{equation}
is the predictive probability according to the model $p$, 
with the probabilities $p(\rho_n=h \mid \rhomn)$ implied by the
$\mbox{PY}$ prior. 
See Section \ref{sub:hier} for more details.
The first $H$ terms in \eqref{pred} formalize borrowing of strength
across $i$, and only the last term corresponds to a separate
subject-specific $\xib_i$. The earlier case 
possibly implies significant regularization,
allowing, for example, to increase the probability of
patterns that are likely for a large group of subjects.
Major regularization occurs for subjects whose predictive distribution
for $\Rb_i$ differs from the corresponding quantity under the unscaled
reference model $\g$.

In anticipation of their use in the following chapters, we introduce explicit notation for the probabilities for a behavior, marginalized by the other subject's parameters. 
Let $\gl \equiv \g(\Rb_n=\rell)$ and
$\pl \equiv p(\Rb_i=\rell\mid\xibm, \rhom)$.
Note that exchangeability among subjects allows  $\pl$ to be deterministically evaluated following Equation \eqref{pred}. 
We discuss more details in Section \ref{sec:priorelicitation}. 

\schwarz
\section{Prior Elicitation}\label{sec:priorelicitation}
We discuss prior choices used in both, the application in Section
\ref{sec:Application} and, with minor adjustments, the simulations in
Section \ref{sec:sim}.  Subsection \ref{sub:hier} briefly discusses
the hyperparameter choices for the hierarchical model, focusing on the
choice and interpretation of $\Rb_i$.
Section \ref{sub:SUN} discusses prior elicitation of hyperparameters
for the unscaled reference models.

\subsection{The Reference Model -- SUN}\label{sub:SUN}
We first introduce the reference model $\g$. 
The model $\g$ is a dynamic state-space model structured to ensure
the desired flexibility.
This class of models is defined by a state equation (or evolution, or
system equation) and an observation equation (or sampling model), and
additional priors. 
For each subject, we define a reference model with prior
$$
\g(\tb_{i,0},\dots,\tb_{i,T_i})=\g(\tb_{i,0})\ \prod_{t=1}^{T_i}
\g(\tb_{i,t}\mid \tb_{i,t-1})
$$
including evolution equation $\g(\tb_{i,t}\mid \tb_{i,t-1})$ and a
prior for the initial state, 
and an observation equation 
$$
\Pr(y_{i,t}=1\mid h_{i,t}) = \Phi^{-1}(h_{i,t}) 
$$
given the state, with $h_{i,t}$ being a linear function of the state
vector $\tb_{i,t}$ (see below). 
While only the prior $\g(\tb_i)$ on the state parameters features in
the inference model \eqref{eq:p},  complementing the prior with
an observation equation for the binary time-series
$\yb_i$ implies a complete Bayesian framework for the marginal
inference on patients (which we shall use later as a a useful device in posterior
simulation). 
The details of the evolution model $g(\tb_{i,t} \mid \tb_{i,t-1})$ and
the linear predictor $h_{i,t}$ are
\begin{align}
  \tb_{i,t}
  &=\Gm\tb_{i,t-1} +\vepsb_{i,t-1} ~~~~  \text{(state
    equation)}  \\ 
 h_{i,t} 
  &=m_i+\zb^\top_{i,t}\tb_{i,t} ~~~~~ \text{ (observation equation)}
\label{eq:g}
\end{align}
with $\tb_{i,0}\sim\N(\zerob,S_0)$ and
$\vepsb_{i,t}\sim\N(\zerob,W_{i,t})$, that implies
$\g(\tb_{i,t}\mid\tb_{i,t-1})$.   
The scalar $m_i$ is a fixed, subject-specific intercept.
Including the fixed intercept in the construction, we can later use
$m_i = \xb_i^\top\deltab$ to achieve matching likelihood factors
under the reference model $\g$ and the inference model $p$. This will
be useful for constructing computation-efficient transition probabilities
in posterior Markov chain Monte Carlo simulation. 
The covariate vectors $\zb_{i,t}$ are subject- and time-specific and may
vary across days, including, for example, indicators for treatment
assignments.
 As usual in dynamic state-space models \citep{petris2009dynamic}, the design matrices $\Gm$ represent the deterministic evolution of the
daily latent state for the seizure probability.
Without loss of generality, we assume that $\Gm$ and $\Wm$ are
shared across all subjects and times.   
The only exception is the
following. 
For each subject, we allow increased evolution noise in the latent states
when known external events happen, such as a 
change of treatment.   
Let $\TT^\star_i$ represent the time index sets marking treatment changes for subject $i$.
For instance, if a subject alters treatment after one month and again after the first year, then $\TT^\star_i=\{1,\,31,\,366\}$).
We use $\Gm=G$ and $\Wm=W$ for all $t\not\in \TT^\star_i$  and
$\Gm=G^\star$ and $\Wm=S_0$ for all $t\in \TT_i^\star$ (see the
Supplementary Material for specific values used in the implementation). 
The first time-varying covariate $z_{i,t,1}$
is defined as the number of days under the active treatment for that
patient. Previous example-subject will have $z_{i,1,1}=z_{i,31,1}=z_{i,366,1}=0,\;z_{i,2,1}=z_{i,32,1}=z_{i,367,1}=1,\;\dots$,  implying that $\th_{i,t,1}$ can be interpreted as treatment reactivity (other, non-active treatments do not affect the likelihood).
The remaining entries, $\th_{i,t,j}$, with $j>1$, capture the autoregressive
component of epilepsy progression and a dynamic intercept with the
corresponding elements $z_{i,t,j}=1$.
Note that both, the number of days under observation $T_i$ and the
sets  $\TT^\star_i$, vary between subjects, implying that the unscaled models are
subject-specific (and therefore necessitates indexed $i$ in $g_i$).

\subsection{Definition of the Events of Interest for HEP data}
\label{sub:hier}

\begin{figure}
    \centering
    \includegraphics[width=\linewidth]{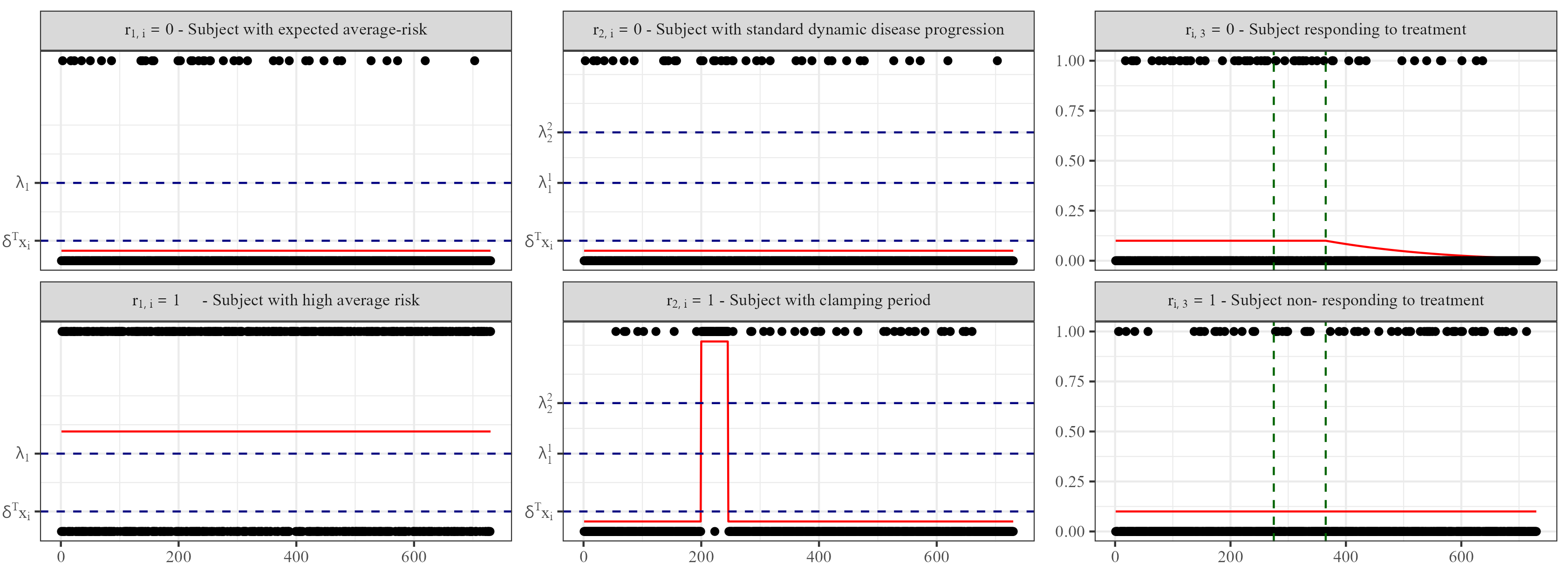}
    \caption{Examples of subjects and latent states. Continuous lines correspond to examples of latent states that imply realization for $r_{k,i}$ (with $k=1,2,3$). 
    In particular, the bottom panels show the realizations of behaviors of interest (i.e., $r_{k,i}=1)$.
    Horizontal lines refer to thresholds of interest for the definitions of $r_{1,i}$ and $r_{2,i}$.
    Changes of treatments, required for $r_{3,i}$,
correspond to the dotted green lines.
Black dots correspond to the binary simulated data (no seizure when drawn below
the curves, seizure when drawn above them.
Detail definitions of $S(\thb_i)$ are given in Section B of the Online Supplementary Material. }
    \label{Example_pattern}
\end{figure}

Prior model $\Pxig$ is completed by introducing the patterns of interest
that define the events $\AAA_\ell$ and the corresponding 
indicators $\Rb_i$. 
We use three basic events of clinical interest
to define  $S(\thb_i)=(r_{1,i},r_{2,i},r_{3,i})^\top\in\{0,1\}^3$, implying
$\Rb_i\in\{0,\dots,7\}$.  
These basic events  are defined in terms of the dynamic
component $\gamma_{i,t} \equiv \zb_{i,t}^\top\tb_{i,t}$
of the probit score.
The first summary statistic, $r_{1, i}$  aims to identify subjects whose
dynamic random effects consistently represent
higher risk,  relative to the reference level $\mu_i=\xb_i^\top\deltab$.
The second summary, $r_{2, i}$, is similar to   $r_{1, i}$, but  focuses on local
slices of the time series and formalizes a conditional probability of high
risk over a local slice given the patient is ``at risk'' 
(see the Supplementary Material for a precise definition).
The indicator $r_{2, i}$ is intended to represent  the notion
of clumping. 
Recall that clumping is defined as periods of frequent seizures prompted by an initial triggering seizure. The associated latent state requires a sudden shift in the time-series, which persists for a significant duration.
Such evolution is uncommon \textit{a priori} under the reference model. 
We define $r_{2,i}$ to indicate that the subject's high-risk time slices likely align with very extreme-risk intervals to enhance model flexibility to account for clamping periods. 
Indicator $r_{3, i}$ compares the 90 days before the last treatment
change with the  most recent part  of the time series, i.e.,
the part under the last active treatment. It aims to capture the
evidence for being a non-responder to the current treatment,
that implies identifying subjects that might require a change of
treatment.
The patterns $R_i$ correspond to combinations of these events, 
to match questions of clinical interest. 
For example, evidence of many subjects having a high probability for
$\{r_{2,i} = 1,r_{3,i}=0\}$ relates to the possibility of correctly
classifying subjects for whom treatment had an effect.
Figure \ref{Example_pattern} shows synthetic examples of data and latent space that ideally corresponding to the described behaviors.
Detail definitions of $S(\thb_i)$, tailored for HEP analysis, are given in Section B of the Online Supplementary Material. 
Any alternative choice of clinically meaningful events (in terms
of the $\thb_i$) could be used.
Also, $R_i$ need not be based on combinations of basic events (like
the $r_{j,i}$ here).  
Any definition with few to moderately many (relative to $n$) possible outcomes for $R_i$ could be used. 
Finally, in the upcoming discussion recall that model
 \eqref{eq:py} and \eqref{eq:p} clusters patients by the probabilities
for $R_i$, not by $R_i$ itself. That is because we do not expect
patterns to be exclusive to subpopulations, but they might be more or less
prevalent in different subpopulations.

\subsection{Hyperparameters elicitation for the Nonparametric Prior.}

Finally we discusthe $\mbox{BNP}$ prior for $P$ in \eqref{eq:p}. 
This prior implies the cluster arrangement of patients and formalizes the borrowing of strength.
We use a Pitman-Yor ($\PY$) process
\citep{pitman1997two}.
The $\PY$ prior is a discrete stochastic process that generalizes the
widely used Dirichlet process (see, for example, 
\cite{Ghoshal:10} for a brief review).  
The $\PY$ process is characterized by a
centering measure $G_0$, a concentration parameter $M$, and a discount
parameter $\sigma$.
Random partitions implied by a $\PY$ prior are similar to the Chinese
restaurant process (implied by the ties under sampling from a Dirichlet process
random measure),  but the $\PY$ process allows
an additional layer of flexibility given by parameter $\sigma$,
$M>-\sigma$. 
These parameters control the degree of clustering in the data. 
The most common parametrization uses $\sigma\in(0,1)$.
In contrast, we suggest and use $\sigma<0$, which requires $M=-k\cdot\sigma$ for
$k$ to take an integer value.
This prior specification implies a maximum number of species and favors a strong borrowing strength while preserving a preference for some large clusters. 
The latter is a desirable feature in our application.
In particular, we use $M=10$ and $\sigma=-1$, implying an upper bound of at most 10 clusters, 
which is close to the cardinality
of the support for $\Rb_i$.
For more details we refer to
\cite{deBlasi2013gibbs}, \cite{bissiri2013species},
\cite{lee2013defining} or
\cite{lijoi2020pitman}.

Finally, we need to specify $G_0$, which is 
the prior for the atoms of the PY random measure in \eqref{eq:py}, and
thus becomes the prior of the unique values $\xibs_h$.
For interpretability it is desireable that
the $\xibs_h$ be informative on the patterns of interest.
We therefor choose $G_0$ to favor weighs $\xib_i$ in the corners
of he simplex. 
For both simulations and real data, we use $G_0=\mathcal{D}ir(1/20, \ldots, 1/20)$.

\section{Posterior Inference}\label{sec:inference}
\subsection{Posterior Sampling} \label{HEP:4.2}

We implement inference using Markov chain Monte Carlo
(MCMC) posterior simulation. 
We use a Gibbs sampling MCMC algorithm, with detail transition
probabilities given in Section A of the Online Supplementary Material. 
In this section, we briefly outline the key elements
of the proposed MCMC scheme including Gibbs and
Metropolis-Hastings (MH) transition probabilities.

An important element of the algorithm is a transition probability to update individual-specific parameters. 
Noting that strong posterior dependence of $\Rb_i$, $\tb_i$, and $\rho_i$ would likely lead to slowly mixing Markov chains under transition probabilities that update only one parameter at a time, we use a $\mbox{MH}$ transition probability to jointly update all subject-specific parameters, targeting  
$p(\tb_i, \Rb_i=\rell, \rho_i=h\mid  \deltab ,\xibs_{k}, \rhob_{\text{-}i}, \yb_i)$. 
Denote by $\tbt_i,\rhot_i,\Rbt_i$ the proposed new values. 
We use a   hierarchical proposal distribution $Q$, composed of
a proposal for $(\tb_i,\Rb_i)$, and a proposal for $\rho_i$
conditional on $\Rbt_i$.
For the earlier, we use posterior inference under
the reference model $g_i$. 
For the latter, we use the implied
conditional under the inference model \eqref{eq:py} and \eqref{eq:p}.
Hence: 
\begin{equation}
\begin{split}
Q(\tbt_i,\Rbt_i=\rellp,\rhot_i=h'\mid \deltab, \yb_i)&\propto  p(\rhot_i=h' \mid \Rbt_i, \xibm,\rhom)\times \g(\tbt_i, \Rbt_i\mid\yb_i, m_i) \\
&\propto p(\rhot_i=h' \mid \Rbt_i, \xibm,\rhom)\times\g(\yb_i \mid \tbt_i, m_i)\ \g(\tbt_i \mid \Rbt_i) \ \g(\Rbt_i=\rellp).  \\
\end{split}\label{ProposalPost}
\end{equation}
The factor uses verbatim the probabilities under the inference
model $p$.  
The proposal for $\tb_i$ (which implies also $\Rbt_i$) needs 
more attention.
% The construction is motivated by considering the hierarchical
% structure of \eqref{eq:p}. Also, not that it is
We use posterior simulation under the {\it unscaled} reference model $\g$.
Recall the definition of model $\g$ in Section \ref{sub:SUN}.
\cite{durante2019conjugate} introduced a computation efficient approach
for posterior sampling of the probit parameters under $\g$, including
in particular a conjugacy result.  
However, the latter requires sampling from a high-dimensional multivariate truncated
normal, which is impractical for large $T_i$. 
Alternatively,  \cite{fasano2021closed} introduced a
sequential Monte Carlo (SMC) strategy that can be used to approximate $\g(\tb_{i,t} \mid \yb_i)$. The algorithm
reduces the size of the truncated normal that is needed for each sampling
step. 
We propose a modified version of this strategy  to return samples from
$\g(\tb_{i, 1:T_i} \mid 
\yb_i, m_i)$ and to work in the presence of missing data. \ech
 Details of the required modifications are discussed in Section A of the Online Supplementary Material. 
Recall the factorization of the joint model in 
\eqref{eq:thxi}. The joint model and proposal $Q$ in \eqref{ProposalPost}
imply MH acceptance ratio
\begin{equation}
\begin{split}
  \alpha
  & =\min\Bigg\{\ 1,\ \
    \frac{p(\yb_i \mid \deltab, \tbt_i)\times p(\thb_i \mid \Rbt_i)
      \times p(\rhot_i=h' \mid \Rbt_i, \xibm,\rhom)\times
      p(\Rbt_i=\rellp\mid\xibm,\rhom)}
    {p(\yb_i \mid \deltab, \thb_i)\times p(\thb_i \mid \Rb_i) \times
      p(\rho_i=h \mid \Rb_i, \xibm,\rhom)\times
      p(\Rb_i=\rell\mid\xibm,\rhom)} 
    \times \\
  &\hspace{1.3cm}\times\ \frac{
 \g(\yb_i \mid \thb_i)\times \g(\thb_i \mid \Rb_i) \times \g(\Rb_i=\rell) \times p(\rho_i=h \mid \Rb_i, \xibm,\rhom)
  }{
 \g(\yb_i \mid \tbt_i)\times \g(\tbt_i \mid \Rbt_i) \times \g(\Rbt_i=\rellp) \times p(\rhot_i = h'\mid \Rbt_i, \xibm,\rhom)
  }   \Bigg\}=   \\
  &= \min\Bigg\{\ 1,\ \
\frac{p(\yb_i \mid \deltab,\tbt_i)\times \plp}{p(\yb_i \mid \deltab,\tb_i)\times
  \pl}\times\dfrac{g(\yb_i \mid \tb_i)\times \gl}{g(\yb_i \mid \tbt_i)\times
  \glp}\;
%\frac{p(\Rbt_i=\rellp\mid\xibm,\rhom)}
%     {p(\Rb_i=\rell\mid\xibm,\rhom)}
% The red part is pli
\Bigg\}.
\end{split}
\end{equation}
Note that in the last line, we used the previously introduced 
$\pl$ and $\gl$. 
Moreover, it is possible to choose $m_i$ in the
unscaled model \eqref{eq:g} to impose the
same likelihood for the scaled and the unscaled
model,  and further simplify the acceptance ratio,
%Letting $\pl=p(\Rb_i=\rell\mid\xibm,\rhom)$, we are left with \ech
% it si defined after (5)"
\[\alpha=\min\left\{\frac{\plp}{\pl}\times\frac{\gl}{\glp}\right\}.\]
Both quantities in the acceptance ratio are easy to compute as $\pl$
can be evaluated from equation \eqref{pred} and $\gl$ are fixed
quantities that can be evaluated outside the $\mbox{MCMC}$ chain by
prior simulations under the reference model.  
Finally, note that if $\ell=\ell'$ the ratio is 1. 
The interpretation of the ratio is straightforward as it inflates the
transition towards behaviors that are common among other subjects - as
results of the clustering - and behaviors that were rare under the
reference moder - as results of the rescaling. 
See the Supplementary Material for details of the remaining transition
probabilities to update $\xib_i$ and $\deltab$,
following standard MCMC transition probabilities.

\subsection{Reporting inference summarizing the Random Partition}

The main inferential targets are posterior probabilities for
the events of clinical interest for each patient, 
${\Pr(\Rb_i=\rell\mid \DD\ ) = }
\EE[\ \II(\Rb_{i}=\rell)\mid \DD\ ]$,
which can be estimated as the corresponding
empirical frequency in the posterior MCMC samples for each subject.
Posterior inference includes a partition of patients by the
probabilities for these events, which can be summarized by
the  posterior similarity matrix $A$, with entries
$a_{ij} = \Pr(\xib_i=\xib_{i'} \mid\DD\ )$. 
Here, $a_{i,j}$
reports the posterior probabilities for subjects $i$ and $i'$ to
cluster together, and can be approximated as the proportion of
MCMC posterior samples in which two subjects are allocated to the same
cluster.
If desired, point estimates for the latent partition can be obtained
using methods implemented, for example, in the \texttt{R} package
\texttt{salso} (\citealp{dahl2024package}).
Finally, note that, if required for specific inferential needs, 
such as precision medicine, the sampling scheme described in this
section allows the collection of posterior samples for the latent
states $\tb_i$ for each subject.
An important consequence of shared information is that such
inference can be generalized to future subjects through 
posterior predictive probabilities of co-clustering (possibly
even conditional on a partial time series $\yb_i$).

\section{Simulations}\label{sec:sim}
\subsection{Simulations setting \ech}
We validate and assess the proposed approach in a simulation study.
Each of the 100 simulations assumes
two groups of subjects characterized by different time-series behavior
of interest, which the model is designed to capture through
separate clusters.
Additionally, to imply different static characteristics $\xb_i$, we
create two subtypes within each of the two clusters cluster, slightly modifying the latent states that represents the simulation truth.
Each of the two simulated clusters contains 50 subjects, both with 25
people for each of the subtypes of time-constant characteristics, implying a total
sample size of 100 individuals.  For each patient we simulated a
binary time series, assuming follow-up over two years with a treatment
change at the end of the first year. The \underline{first simulated cluster}
are patients who did not exhibit elevated risk and are
treatment responders, corresponding to intended
(intended since the simulation truth is outside the inference model)
$S(\thb_i)=(r_{1,i},r_{2,i},r_{3,i})^\top 
= (0,0,0)^\top$,  i.e. implying simulation truth
$\Rb_{i} = 0$ in our model.
The \underline{second cluster} are patients who (like cluster 1
patients)
are treatment responders, and have no elevated average risk, but who
do experience clumping (i.e., $r_{2,i}=1$, implying simulation truth
$\Rb_i= 2$). 
With all patients under this scenario being treatment responders the model should report this feature for all the simulated patients. 
Finally, we also generate missingness indicators. 
In the first cluster, marginal inference is challenging due to the low probability for seizures, which makes it likely that determining whether the treatment change was beneficial relies on few events, both before and after the change of treatment. 
In the second group, the inference is even more complex since the presence of a clamping period in the second window of the time series can easily mask an ineffective treatment, and we aim to predict behavior that is rare \textit{a priori}.
A detailed explanation of the data-generating process and representation of the latent states is available as Online Supplementary Material, in Section C. 

The purpose of this simulation is to
assess the proposed inference in comparison to alternative
marginal approaches demonstrating the effectiveness of the borrowing of information approach.
We compare our strategy with a two step approach, using
first a point estimate $\deltah$ based on a 
(static) probit regression parameters
$\Pr(y_{i,t}=1)\equiv\Phi^{-1}(\xb_i^\top\deltab)$, 
and separate instances of the $\g$ model with fixed $m_i=\xb_i^\top\deltah$,
for the inference on subject-specific summaries, including in
particular posterior probabilities for $\Rb_i$. 
This two-step approach retains the flexibility of the dynamic probit
model but shares information only through an Empirical Bayes ($\mbox{EB}$) estimates for
$\deltah$. 
We compare the two models by reporting multi-class cross-entropy.
Note that the described simulation scenario implies a deterministic $\Rb_i = \ell^\star_i$, so that the multi-class cross-entropy is equivalent to the average of the $\log_2$ posterior probability of the simulation, evaluated as
\[ H = \frac{1}{N}\sum_{i=1}^N\log_2\{\ p(\Rb_i=\ell^\star_i\mid\DD)\
  \}.\]
%with $\ell^\star_i$ denoting the simulation truth on $\Rb_i$. %\bchGP The optimal simulations result is in 0.\ech
According to this metric, an ideal model would result in $H = 0$ for each simulation. 

\subsection{Simulations Results}
\begin{figure}
    \centering
    \includegraphics[width=.8\linewidth]{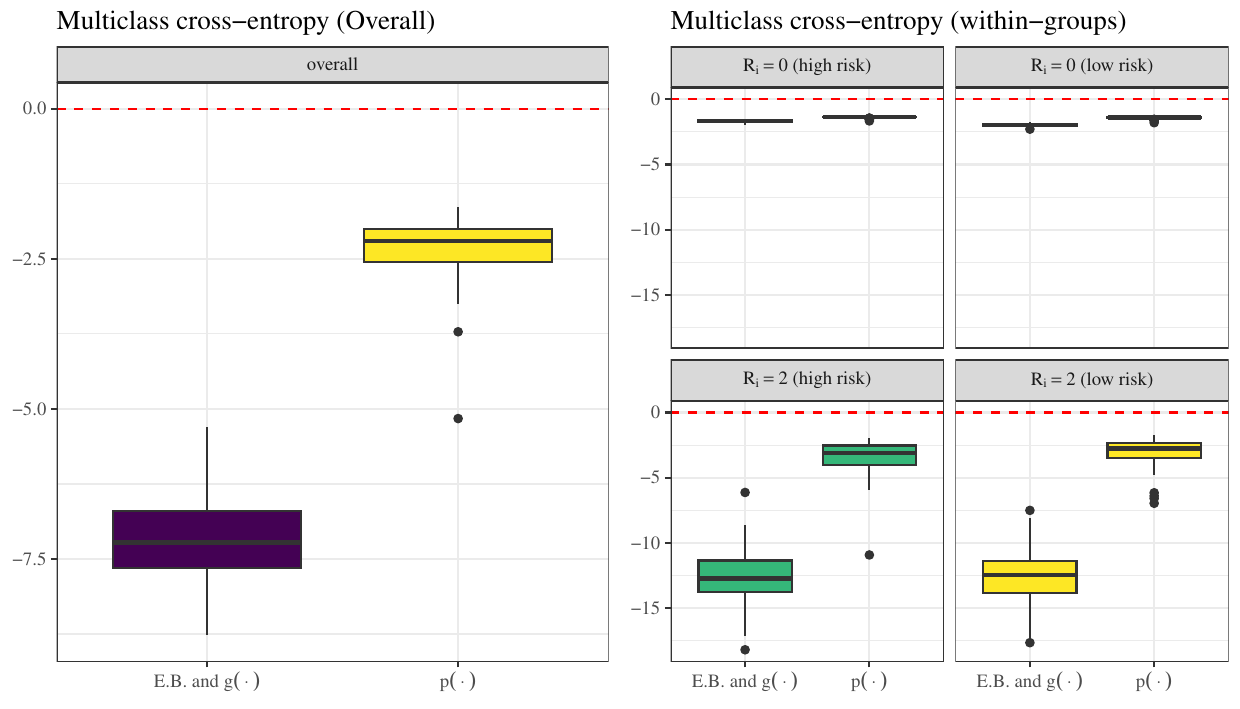}
    \caption{Simulation study.
      Boxplots of log posterior probability for $R_i$ 
      over 50 simulations.
      Here $p(\cdot)$ refers to inference under the proposed model and
      $g(\cdot)$ indicates the reference model. The four boxplots in
      the right panel are stratified by the simulation truth and by
      the two subtypes of high versus low risk in the simulation
      truth. }
    \label{fig:sim-res}
\end{figure}

The results of the simulations are summarized in Figure
\ref{fig:sim-res}.  The figure shows a clear increase in
posterior probability of the simulation truth
under the proposed inference approach
versus the same under the reference
model, using separate inference for each patient. 
As seen in the right panel of Figure \ref{fig:sim-res}
the relative gain in log posterior probability varies across the
subgroups, being most notable for clumping subjects ($\Rb_i=2$).
This might be due to the impact of information sharing being
higher for patients with patterns that are {\it a priori} less likely
under the unscaled model. 
The $\SUN$ prior in the reference model can over-regularize rare
(\textit{a priori}) clumping behavior.  

\section{Analysis of the HEP Data}\label{sec:Application}
For analysis of the $\HEP$ data, we collected 2500  
posterior samples using five parallel $\mbox{MCMC}$ chains.  
Each chain was run for 13.500 iterations, with a burn-in period of
1000 and a thinning interval of 25 iterations. 
$\MCMC$ simulations were initialized as follows. 
First, $\deltab$ was initialized with maximum likelihood
estimates under a static probit model. 
Conditioning on the initial value of $\deltab$, we used the posterior
expected values implied by the unscaled model to initialize the latent
space parameters for each
subject i.e., $\EE[\thb_i\mid\yb_i]$ under the reference
model $\g$ in \eqref{eq:g}, estimated  separately for each
patient.
Initial values for $\Rb_i$ were initialized as implied by the
initial values for $\thb_i$.
The nonparametric components of the model were initialized using
homogeneous clusters constructed according to current values
of $\Rb_i$ and the expected value implied for the vector
$\xib_h^\star$.   

\begin{figure}
\centering
\begin{minipage}{12.3 cm}
\rotatebox{180}{\includegraphics[width=\textwidth]{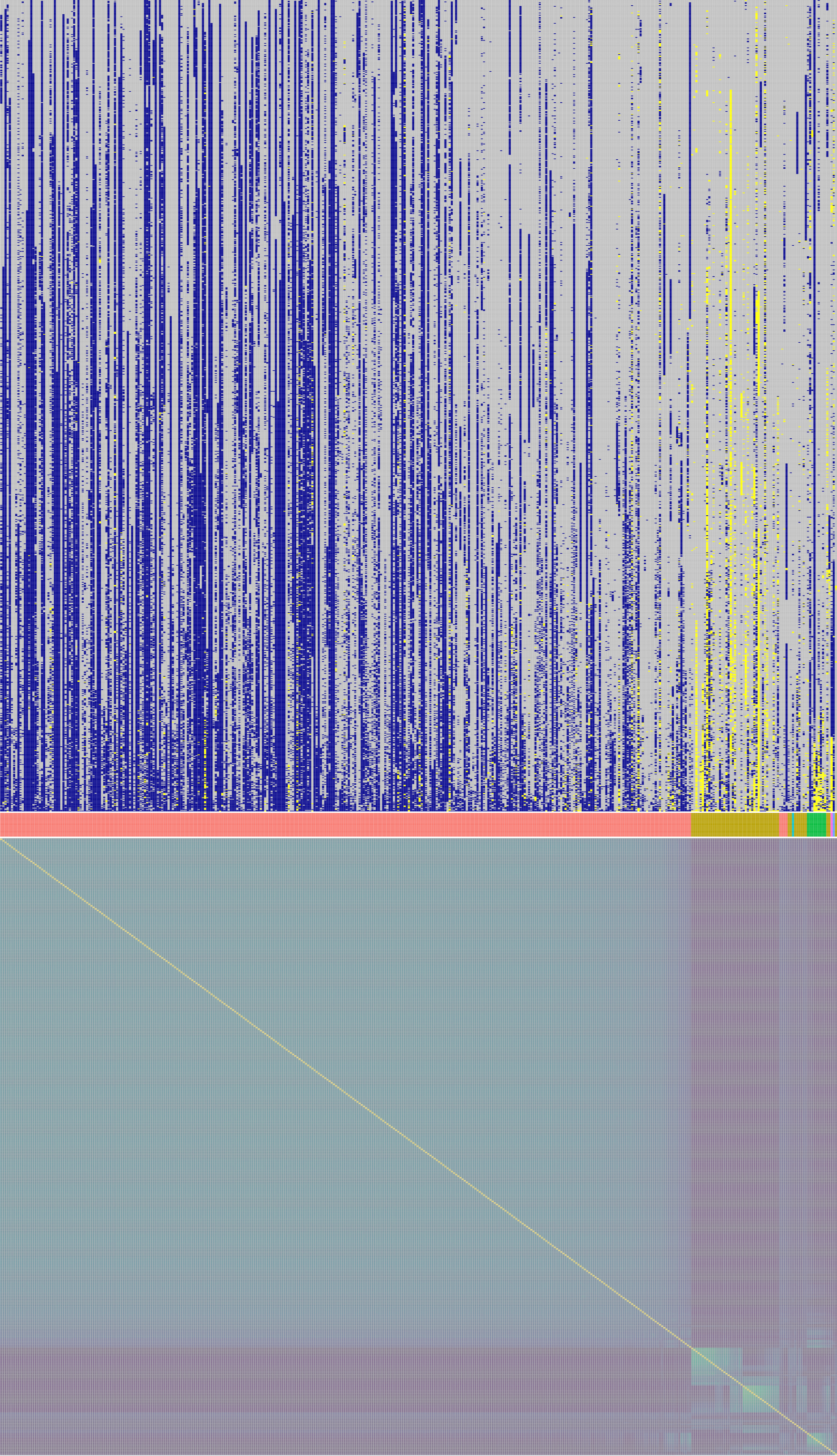}}
\end{minipage}%
\begin{minipage}{3 cm} % Adjust width for the legend column
\includegraphics[width=3cm]{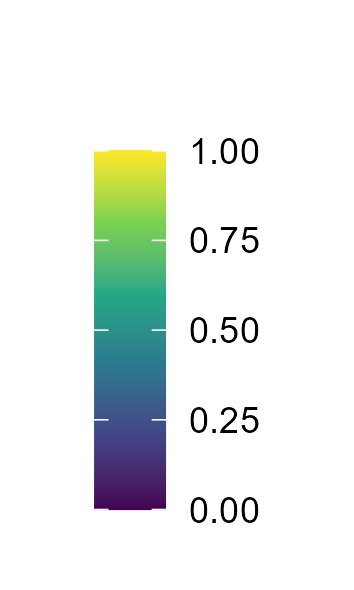}\\[0.5cm]
  \hspace*{0.35cm}\includegraphics[width=3cm]{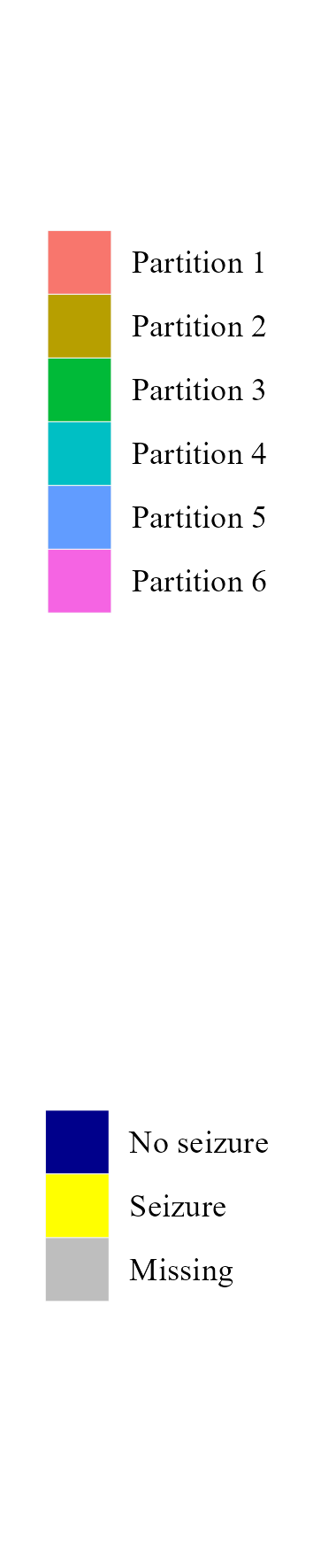}
\end{minipage}
\caption{Summary of the posterior estimated partition of  the
  $N=390$  patients in the $\HEP$ data.    
  The heatmap in the top panel shows the posterior co-clustering
  probabilities, i.e., the $N \times N$ matrix $A$ of co-clustering
  probabilities $a_{ij}=\Pr(\xib_i=\xib_j \mid \DD)$.  
  The bottom panel shows the observed data with $N$ columns for
  the patients and $T_i$ rows for repeat visits for the $i$-th patient
  (with grey filling in beyond row $T_i$).
  The colored bars dividing the two plots represent a point estimate for the partition (under the $\mbox{VI}$ loss).}
\label{fig:HEP_results}
\end{figure}

\underline{Partition:}
The population results of the hierarchical model on HEP data are summarized in Figure \ref{fig:HEP_results} and Table \ref{tab:HEP}.
Figure \ref{fig:HEP_results} shows the partition of patients
that is obtained by minimizing the variation information ($\mbox{VI}$) loss function.
Table \ref{tab:HEP} reports posterior
probabilities  for $\Rb_i$, that is, for the clinical events of
interest, for each of the clusters under the 
estimated partition $\hat\varrho_i$. 
The most common behavior is $\Rb_i=0$ i.e., non-clumping subjects that
are treatment-responsive and with relative-low risk.  
These subjects are 
marked with red in the partition bar in the center of Figure
\ref{fig:HEP_results}, and are labeled as \textit{Partition 1} (in the
Figure and in Table \ref{tab:HEP}).
\textit{Partition 3} is likely associated with clumping subjects that
are treatment responders.  Subjects assigned to  \textit{Partition 2} and
the singleton clusters 4, 5 and 6 
show evidence of non-response to treatment or higher risk levels with
respect to the reference level set by $\xb_{i}^\top \deltab$.

\begin{table}[bt]
\centering
\resizebox{\textwidth}{!}{
\begin{tabular}{lrrrrrrrrr}
\toprule
&\multicolumn{8}{c}{$\Pr(\Rb_i=r_\ell\mid\hat\varrho_i=h,\ \mathcal{D})$} & \\
  \midrule
$(r_{i,1},\,r_{i,2},\,r_{i,3})$ &
\multicolumn{1}{c}{$(0,\,0,\,0)$} &
\multicolumn{1}{c}{$(0,\,0,\,1)$} &
\multicolumn{1}{c}{$(0,\,1,\,0)$} &
\multicolumn{1}{c}{$(0,\,1,\,1)$} &
\multicolumn{1}{c}{$(1,\,0,\,0)$} &
\multicolumn{1}{c}{$(1,\,0,\,1)$} &
\multicolumn{1}{c}{$(1,\,1,\,0)$} &
\multicolumn{1}{c}{$(1,\,1,\,1)$}\\
    \midrule
$\hat\varrho_i=h$ &
\multicolumn{1}{c}{$\Rb_i=0$} &
\multicolumn{1}{c}{$\Rb_i=1$} & 
\multicolumn{1}{c}{$\Rb_i=2$} & 
\multicolumn{1}{c}{$\Rb_i=3$} &
\multicolumn{1}{c}{$\Rb_i=4$} & 
\multicolumn{1}{c}{$\Rb_i=5$} &
\multicolumn{1}{c}{$\Rb_i=6$} & 
\multicolumn{1}{c}{$\Rb_i=7$} & Size \\ 
  \midrule
Partition 1 & 0.94 & 0.01 & 0.03 & 0.01 &   &  &  & & 326  \\ 
Partition 2 & 0.03 &  & 0.05 & 0.03 & 0.01 & 0.02 & 0.43 & 0.43 & 52 \\ 
Partition 3 & 0.05 & & 0.87 & 0.01 & 0.01 & & 0.06 & & 9 \\ 
Partition 4 & 0.15 & 0.31 & 0.01 & 0.20 & & 0.05 & & 0.27 & 1 \\ 
Partition 5 &  & 0.01 & 0.07 & 0.91 &  & &  & & 1 \\ 
Partition 6 & 0.03 & 0.04 & 0.39 & 0.55 &  & & & & 1 \\ 
   \bottomrule
\end{tabular}}
\caption{Posterior probability for the latent state of the time series to manifest a given behavior. The probabilities are evaluated within the partition of the data minimizing binder loss function (i.e., Rows sum to 1). The last column shows the size of the partition.}\label{tab:HEP}
\end{table}

\begin{figure}
\begin{minipage}{\textwidth}

\end{minipage}
    \centering
    \includegraphics[width =\linewidth]{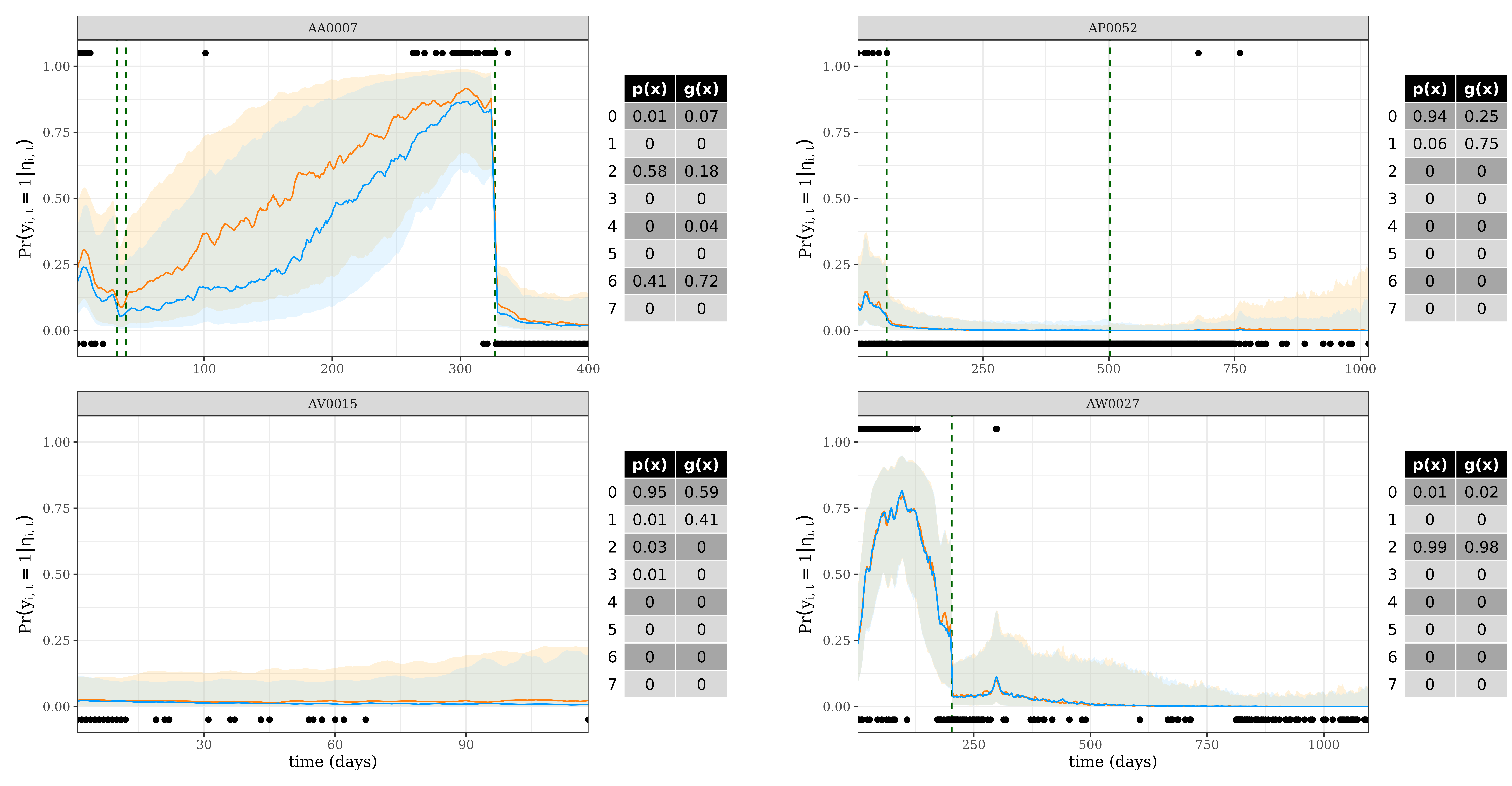}
    \caption{
      Fitted mean curves under
      $p$ (blue, lower curve) and $\g$ (orange, upper curve).
      Changes of treatments correspond to the dotted green lines. Black dots correspond to the binary observed data (no seizure when drawn below the curves, seizure when drawn above them). The absence of such points indicates missing data. 
      The tables show posterior probabilities for $\Rb_i=\ell$ under the
      two models. }
    \label{fig:marginal}
\end{figure}

\underline{Patients.}
We complement the population analysis with some details on the
marginal inference for some patients. 
In Figure \ref{fig:marginal}, we summarize inference for
four patients all of whom reported periods of missing
data.  For each subject, smoothed credible intervals of the
probability of a seizure event are shown for inference under model 
$p$ and under the unscaled model $\g$, respectively.
Changes in treatment (if any) are indicated as vertical dotted lines.
The table on the side of each figure shows the 
posterior probabilities for $R_i=\ell$ under the two models.  
Under $p$ the latter are based on data from all patients
in the same cluster, whereas there is no such notion of pooling under
$\g$. 
Results concerning marginal inference for all other subjects
in the sample are provided in the Supplemenatry Material.
For patient \ALB0007, inference under $p$ includes 
a high posterior probability for both, $R_i = 2$ and $R_i = 6$,
indicating that the individual is a treatment responder
subject with clumping behavior. 
The data point on day $t=100$ strongly influences inference under
$\g$, leading to fast growing fitted probabilities $\Pr(y_{it}=1)$.
In contrast, inference under $p$ is less sensitive to the single
observation, and shows more smoothing, but eventually, after more
observations with $y_{i,t}=1$, it ends up fitting equally high values.
Patients \NYU0052 and \UAB0015 show high posterior probabilities for
$R_i=0$, i.e.,
low overall seizure risk, no clumping and treatment responder.
In contrast, for patient \UCSF0027, we find strong evidence for $\Rb_i=2$, leaving the borrowing of information
within clusters less crucial. Both models provide very similar 
posterior probabilities for $\Rb_i$.  Overall, 
in the presence of high levels of noise, borrowing of information
allows more robust inference that is less sensitive to sporadic
outliers.
The explicit inclusion of $\Rb_i$ in the construction of the
hierarchical model favors posterior inference to focus on trajectories
with meaningful clinical interpretations.

Finally, we note that the proposed model implements
borrowing of strength across patients on the pre-defined events of
clinical interest coded by $\Rb_i$, while remaining unconstrained about the
fitted profiles beyond these events. 
This approach enables the flexible modeling of details of the time series
for which a joint model would be impossible or impractical.
If desired, more population level summaries can be inferred
by suitable summaries of the posterior Monte Carlo sample, 
using posterior post-processing similar to \citealp{wang2022bayesian}. 
For example, one could use this approach to summarize evidence about
treatment effectiveness. Related results are shown in Section C of the Online Supplementary Material. 
Such analyses should, however, be interpreted with caution.  
Formal analyses would require the specification of
suitable pre-defined $\Rb_i$ and/or more structural assumptions in the
model.

\section{Conclusions}\label{sec:conclusion}
We introduced a novel Bayesian semi-parametric approach to effectively
model un-aligned longitudinal binary time-series data.  The proposed inference approach is
applicable for any data with binary time series that are
unaligned in time, but it was specifically motivated by a data set of
indicators for daily epilepsy seizure occurrences.  Our strategy
demonstrates how borrowing of strength between time series can be
achieved even without synchronization in time.  The construction
provides a flexible framework that accommodates heterogeneity across
patients and time, while leveraging commonalities across
individuals. The latter are defined in terms of a set of pre-defined
events of clinical interest (patterns in the longitudinal data).
This approach facilitates the discovery of homogeneous (with respect to these
patterns) subgroups within the patient
population.

The model sets up a hierarchical structure to cluster patients with
respect to shared probabilities for the pre-defined patterns. 
This allows to share strength across patients who exhibit similar seizure patterns, a
feature that proves advantageous in the analysis of complex
longitudinal data with misaligned events.
The proposed framework can easily be adapted to other applications that
involve non-aligned time-series data but a notion of shared patterns
across subjects, making it versatile for a range of
medical and epidemiological studies and can also be adapted to different analysis scenarios, where principled models can be elicited marginally for subjects, yet standard hierarchical structures - motivated by inferential propose -  cannot be adopted without compromising marginal models flexibility.   

The limitations of the proposed approach are
mainly related to the extreme flexibility that makes the model
computationally expensive.  Also, the flexibility inherent in the
state space model limits the ability to learn and infer characteristics for future subjects. Learning across subjects only occurs by way of
the probabilities for the pre-defined events of interest,
while other features of the time series remain independent across
subjects.

\begin{comment}
removing -> we just need it for Biometrics  
\subsubsection*{Acknowledgement}
The authors appreciate use of the HEP study data, and acknowledge
advice and feedback from Giovani da Silva and ...

\subsubsection*{Funding}
This research was supported by the National Institutes of Health (NIH)
under grants  5R01NS133040-02.
\end{comment}

\bibliographystyle{apalike}
\bibliography{bib}

\newpage

\appendix

\section{ Algorithms and MCMC Details}\label{apx:A}
We here give more details of the algorithms used for posterior
simulation of the time varying (and patient-specific) parameters in the
probit model. 
First, we briefly review "Algorithm 3" 
of \cite{fasano2021closed}, which we will be denoted as Algorithm 1.  
Algorithm 1 is at the core of our posterior simulation strategy, which
we stated as Algorithm \ref{algo:Gio_2nd}. Algorithm 1
is used as \ech a proposal distribution in the proposed posterior
MCMC algorithm.

For simplicity, we suppress the subscript $i$ for patient.

\subsection{A One-step Lookahead Particle Filter from Fasano et al. (2021)}\label{Apx:FasandGio}

The version of the algorithm we report here is a special case of the
original "Algorithm 3" from \cite{fasano2021closed}, that focuses on
univariate time-series and uses $k=0$. 
In the original algorithm, the parameter $k+1$ represents a
\textit{delay offset}.  
With $k=0$, the sampler reduces to a classical particle filter.
Thus, what we describe here is a one time lookahead particle filter
for a univariate binary time series. 
This algorithm allows to sequentially sample from the marginal
posterior distributions $g(\theta_t \mid \yb)$ for $g$ as introduced
in Section 4.  
For notation purposes, we introduce notation from
\cite{fasano2021closed} and define: 
\begin{itemize}
\item Filters vectors $a_t^{(r)}$ that represent the mean of the
  Gaussian filtering distribution for the multivariate latent state
  $\tb_t$ for each of the $r$ particles.
\item Matrices $P_{t|t-1}$ that represent the covariance matrix of the
  Gaussian filtering distribution for the state vector $\tb_t$ that
  accounts for the uncertainty at time $t-1$.
\item Scalar dispersion parameter $S_{t|t-1}$, computed from the
  observation model combining the observation noise and the uncertainty
  due to the projected filtering state.
\item Scalar $r^{(r)}_{t|t'}$ that represents the mean of the Gaussian
  distribution of the latent variables associated with the filter.
\item Random variables $\zeta_{t|t}^{(r)}$ that is the latent Gaussian
  variable associated with the binary observations $y_t$.  These
  augmented variables are used to compute the updated states in the
  filters.  During the sampler, these random variables are drawn from a
  univariate truncated normal distribution, with supports $
  \mathbb{A}^{|y_t}$.  The notation suggests that the supports are
  implied by observed values of $y_t$.
\end{itemize}
Furthermore, using notation introduced in the main
manuscript, $G_t$ is the state transition matrix, $W_t$ is the process
noise covariance matrix, and $\zb_t$ are the time-varying design
vectors. 
For theoretical details, we refer to the original paper and
\cite{doucet2009tutorial}.  
The sketch of the algorithm is shown as Algorithm \ref{algo:Gio_1st} below.
The algorithm produces marginal samples that can be used to produce
Monte Carlo estimates of the time-series posterior moments.
%Note that this class of algorithm improves both its efficiency and
%accuracy when the time dimension $t$ is large.  
\begin{algorithm}
\caption{One-step Lookahead Particle Filter for sampling from $g(\theta_t \mid \yb)$ \hfill \cite{fasano2021closed}}
\begin{algorithmic}[1]
\Require A binary time series, the number of filters $R$, and model's hyperparameters.
\State Initialize $a^{(r)}_{0|0} = m_0$ for $r = 1, \dots, R$ and $P_{0|0} = S_0$.
\State Sample $\tb^{(1)}_{1|1}, \dots, \tb^{(R)}_{1|1}$ from $\mathcal{N}(a_0,S_0)$.
\For{$t$ from $1$ to $T$}
    \State Set $P_{t|t-1} = G_{t}P_{t-1|t-1}G^{\top}_{t} + W_{t}$.
    \State Set $S_{t|t-1}=\zb_t^\top P_{t|t-1} \zb_t +1$.
    \State Set $P_{t|t} = P_{t|t-1} - P_{t|t-1}\zb_{t}^{\top}S^{-1}_{t|t-1}\zb_{t}P_{t|t-1}$.
    \For{$r = 1, \dots, R$}
        \State Set $a^{(r)}_{t|t-1} = G_{t}a^{(r)}_{t-1|t-1}$.
        \State Set $r^{(r)}_{t|t-1}=\zb_t^\top a^{(r)}_{t|t-1}$.
    \EndFor
    \State Perform resampling $a^{(r)}_{t|t-1}$ and   $r^{(r)}_{t|t-1}$ according to the probit likelihood.
    \For{$r$ from $1$ to $R$}
        \State Sample $\zeta^{(r)}_{t|t}$ from a $\mathcal{TN}(r^{(r)}_{t|t-1}, S_{t|t-1}, \mathbb{A}^{|y_t})$.
        \State Set $a^{(r)}_{t|t} = a^{(r)}_{t|t-1} + P_{t|t-1}\zb^{\top}_{t}S^{-1}_{t|t-1}\big(\zeta^{(r)}_{t|t}- r^{(r)}_{t|t-1}\big)$.
        \State Sample $\tb^{(r)}_{t}$ from a $\mathcal{N}_p\big(a^{(r)}_{t|t}, P_{t|t}\big)$.
    \EndFor
\EndFor
\State \Return $\tb_{1}^{(1)}\ldots\ \tb_{T}^{(1)}\ldots\ \tb_{1}^{(R)}\ldots\tb_{T}^{(R)}$.
\end{algorithmic}\label{algo:Gio_1st}
\end{algorithm}

\subsection{One-step Lookahead Particle Filter for Posterior Latent Trajectories}\label{Apx:OurPF}

Algorithm \ref{algo:Gio_2nd} is used to sample from the proposal distribution defined in Section 5 for the Metropolis-Hastings step.
Compared to Algorithm \ref{algo:Gio_1st}, Algorithm \ref{algo:Gio_2nd} allows sampling from the joint posterior distribution of $(\tb_1, ..., \tb_T)$, allows for missing values among data, and allows for the inclusion of time constant parameters like $\delta$ to be part of the probit model as well.
To sample from the joint posterior, we propagate the filters from our current imputed sample; i.e., we replace $a_{t-1|t-1}$ with $\tb_{t-1}^{(r)}$ and $P_{t-1| t-1}=0$.
Moreover, the need to obtain joint posterior samples implies that the resampling step involves the entire particle trajectories.
This can make Algorithm \ref{algo:Gio_2nd} to be very inefficient as it requires the sampling step to involve parameters from $1,\ldots,t$, but it suits our proposes since our $\mbox{MCMC}$ sampling scheme requires just one marginal posterior sample for iteration. 
Extending this algorithm to serve as a $\mbox{SMC}$ sampling
method for sample size $\approx R$ would likely require to instead
adopt a backward smoothing strategy. 
As most sequential sampling algorithms, Algorithm \ref{algo:Gio_2nd}
can be easily generalized to incomplete data as it is often possible
to propagate uncertainty in parameters proposals and weights without
the need to sample the missing values. 

\begin{algorithm}
\caption{Sampler for posterior trajectories $g(\tb_1,\ldots,\tb_T \mid \yb)$ }
\begin{algorithmic}[1]
\Require A binary time series, the number of filters $R$, and the model's hyperparameters.
\State Initialize $a^{(r)}_{0|0} = m_0$ for $r = 1, \dots, R$ and $P_{0|0} = S_0$.
\State Sample $\tb^{(1)}_{0}, \ldots, \tb^{(R)}_{0}$ from $\mathcal{N}(a_0,S_0)$.
\For{$t$ from $1$ to $T$}
\If{$y_t$ is \textit{missing}}
    \For{$r = 1, \dots, R$}
        \State Sample $\tb^{(r)}_{t}$ from a $\mathcal{N}_p\big(\tb^{(r)}_{t-1},  W_{t}\big)$.
    \EndFor
\Else
    \State Set $P_{t|t-1} = W_{t}$.
    \State Set $S_{t|t-1}=\zb_t^\top P_{t|t-1} \zb_t +1$.
    \State Set $P_{t|t} = P_{t|t-1} - P_{t|t-1}\zb_{t}^{\top}S^{-1}_{t|t-1}\zb_{t}P_{t|t-1}$.
    \For{$r = 1, \dots, R$}
        \State Set $a^{(r)}_{t|t-1} = G_{t}\tb^{(r)}_{t-1}$.
        \State Set $r^{(r)}_{t|t-1}=\zb_t^\top a^{(r)}_{t|t-1}+m_i$.
    \EndFor
    \State Perform resampling $\tb^{(r)}_{1},\ldots,\tb^{(r)}_{t-1}$ and $r^{(r)}_{t|t-1}$ according to the probit likelihood (i.e. given $\deltab$).
    \For{$r$ from $1$ to $R$}
        \State Sample $\zeta^{(r)}_{t|t}$ from a $\mathcal{TN}(r^{(r)}_{t|t-1}, S_{t|t-1}, \mathbb{A}^{|y_t})$.
        \State Set $a^{(r)}_{t|t} = a^{(r)}_{t|t-1} + P_{t|t-1}\zb^{\top}_{t}S^{-1}_{t|t-1}\big(\zeta^{(r)}_{t|t} - r^{(r)}_{t|t-1} -m_i \big)$.
        \State Sample $\tb^{(r)}_{t}$ from a $\mathcal{N}_p\big(a^{(r)}_{t|t}, P_{t|t}\big)$.
    \EndFor
    \EndIf
\EndFor
\State Sample $r^\star\in\{1,\ldots,R\}$.
\State \Return $\tb_{1}^{(r^\star)},\ldots,\tb_{T}^{(r^\star)}$.
\end{algorithmic}
\label{algo:Gio_2nd}
\end{algorithm}

\subsection{Gibbs sampler transition probabilities}\label{Gibbs} 
We provide details for the transition probabilities
used for posterior MCMC simulation.

\subsection*{Update of $\tb_i,\Rb_i,\rho_i$}

We first recall notation. Let
$\{\xibs_1,\ldots,\xibs_H\}$ denote the unique values among
$\{\xib_1,\ldots,\xib_n\}$, 
with $H$ denoting the number of unique
values.
Let $n_h$ denote the multiplicity of $\xibs_h$.
Also, recall that $\rho_i=h$ if $\xib_i = \xibs_h$.
Similarly let $\xib\sm_h$, $n\mm_h$ and $H^-$ denote the same for
$\{\xib_\ell;\; \ell \ne i\}$. 

Here we discuss in detail the Metropolis-Hastings step required to
sample subject specific parameters.
\begin{itemize}
  \item {Target distribution:}
\begin{equation*}
\begin{split}
  p(\tb_i,  \rho_i=h\mid \yb_i,\deltab,\ {\rhobm},\xibm) \propto\
  & p(\yb_i \mid \deltab,\tb_i)\ 
    p( \tb_i\mid \rho_i=h,\ {\rhobm},\xibm)\ p(\rho_i=h\mid {\rhobm},\xibm)\\
=\ & p(\yb_i \mid \deltab,\tb_i)\
    p( \tb_i, \Rb_i=\ell \mid \rho_i=h,\ {\rhobm} ,\xibm)\ p(\rho_i=h\mid {\rhobm},\xibm)\\
 =\ &
p(\yb_i \mid \deltab,\tb_i)\
    p( \tb_i\mid \Rb_i=\ell)\,
    %\underbrace{p(\yb_i \mid \deltab,\tb_i)\ p( \tb_i\mid\boldsymbol{R_i}=r_\ell)\ }_{\text{Proposed with the $g$}}
 %\underbrace{{p(\Rb_i=r_\ell \mid \xibs_k,  \rho_i=k) \ p(\rho_i=k\mid \rhobm)}}_{\text{Proposed with the full conditioned}}\\
    {p(\Rb_i=\ell \mid \rho_i=h,\ {\rhobm},\xibm) \ p(\rho_i=h\mid {\rhobm}, \xibm)}
    \end{split}
\end{equation*}
The first  two factors, by the construction of the hierarchical model, are equal to:
\begin{equation}
  p(\yb_i \mid \deltab,\tb_i)\ \g( \tb_i \mid \Rb_i = \ell,\ m_i=\mu_i), \label{like}
\end{equation}
with $\mu_i= \xb_i^\top\deltab$,  using $\deltab$  fixed at the 
currently imputed value.
That is, $m_i$ is fixed by conditioning on $\deltab$.
The last two factors
can be alternatively factored as
$$
p(\rho_i = h\mid \Rb_i=\ell, {\rhobm}, \xibm)\,
\Pr(R_i=\ell \mid  \rhobm,\xibm),
$$
with 
% Without loss of generality assume $i=n$.  
\begin{equation}
  p(\rho_i = h\mid \Rb_i=\rell,\ \rhobm, {\xibm}) =
  \begin{dcases}
    \kappa_{i,\ell}\times \frac{ n\mm_h - \sigma}{N+M-1} \times \xi^\star_{h,\ell} & if\ h=1,\dots,H\mm\\
    \kappa_{i,\ell}\times \frac{M+\sigma\cdot H\mm } 
  {M+N-1}\times \frac{a_\ell}{\sum_{r=1}^L a_r} \hspace{1cm}& if\ h=H\mm+1.\\
\end{dcases}\label{SSM_art}
\end{equation}
with
$1/\kappa_{i,\ell} = p(\Rb_i=\rell\mid\xibm,\rhobm)$.
\item {\em Proposal distribution $Q$:}
 Let again $\g$ be the unscaled model, conditioning to $m_i=\mu_i=\xb_i^\top \deltab$ and $h(\rho_i \mid \cdot)$ denote the conditional probability in 
\eqref{SSM_art}, with $\Rb_i=\Rbt_i$.  
% see as an example \citealp{neal2000markov}. 
% 
%\ref{eq:rho} can be used, conditioning to $\Rb_i=r_\ell$, for proposing.
We propose a new values for $\tilde\tb_i$ and $\tilde\rho_i=h$ using 
the hierarchical \textit{proposal} distribution $Q(\cdot)$:
\begin{equation}
\begin{split}
Q(\tilde\rho_i,\tilde\tb_i,\tilde\Rb_i \mid \yb_i,\deltab)&= h(\tilde\rho_i\mid \Rbt_i  )\ \g(\tilde\tb_i,\Rbt_i \mid \yb_i,\ {m_i}=\mu_i)\\
&=  h(\tilde\rho_i\mid \Rbt_i )\ \g(\tilde\tb_i \mid \yb_i,\ {m_i}=\mu_i).  \\
\end{split}\label{Proposal1}
\end{equation}
We sample from \eqref{Proposal1} in the following way. 
We first use Algorithm \ref{algo:Gio_2nd} to sample from
$\g(\tbt_i \mid \yb_i,\ {m_i}=\mu_i)$, 
and then generate from \eqref{SSM_art} to sample from the first part
conditional on $\Rbt_i=\rellp$ (as implied by the generated $\tbt_i$
generated before). To evaluate the proposal in the acceptance ratio below we further
simplify the second factor in the last expression. 
\begin{equation}
\begin{split}
\g(\tilde\tb_i,\tilde\Rb_i=\rellp \mid \yb_i,\ {m_i}=\mu_i)\propto&\ \g(\yb_i \mid {m_i}=\mu_i,\ \tilde\tb_i)\ \g(\tilde\tb_i,\tilde\Rb_i\mid {m_i}=\mu_i) \\
  \propto&\ \g(\yb_i \mid {m_i}=\mu_i,\ \tilde\tb_i)\ \g(\tilde\tb_i\mid\tilde\Rb_i=\rellp)\ \g(\tilde\Rb_i=\rellp) \\
    \propto&\ \g(\yb_i \mid {m_i}=\mu_i,\ \tilde\tb_i)\ \g(\tilde\tb_i\mid\tilde\Rb_i=\rellp)\ \glp 
\end{split}\label{proposal2}
\end{equation}
With $\gl \equiv \g(\Rb_i=r_\ell\mid {m_i}=\mu_i)$ as  previously defined.  Similarly recall definition of $\pl\equiv p(\Rb_i=\rell\mid\xibm,\rhobm)$. 

\item {\em Acceptance probability:}
  This implies an acceptance probability $\alpha$ that can be evaluated using \ref{like} and \ref{proposal2} as follows:
\begin{equation}
\begin{split}
  \alpha
  & =\min\Bigg\{\ 1,\ \
    \frac{p(\yb_i \mid \deltab, \tbt_i)\times p(\thb_i \mid \Rbt_i)
      \times p(\rhot_i=h' \mid \Rbt_i, \xibm,\rhom)\times
      p(\Rbt_i=\rellp\mid\xibm,\rhom)}
    {p(\yb_i \mid \deltab, \thb_i)\times p(\thb_i \mid \Rb_i) \times
      p(\rho_i=h \mid \Rb_i, \xibm,\rhom)\times
      p(\Rb_i=\rell\mid\xibm,\rhom)} 
    \times \\
  &\hspace{1.3cm}\times\ \frac{
 \g(\yb_i \mid \thb_i,m_i=\mu_i)\times \g(\thb_i \mid \Rb_i) \times \g(\Rb_i=\rell) \times h(\rho_i=h \mid \Rb_i, \xibm,\rhom)
  }{
 \g(\yb_i \mid \tbt_i,m_i=\mu_i)\times \g(\tbt_i \mid \Rbt_i) \times \g(\Rbt_i=\rellp) \times h(\rhot_i = h'\mid \Rbt_i, \xibm,\rhom)
  }   \Bigg\}=   \\
    & =\min\Bigg\{\ 1,\ \
    \frac{p(\yb_i \mid \deltab, \tbt_i)\times p(\thb_i \mid \Rbt_i)
      \times p(\rhot_i=h' \mid \Rbt_i, \xibm,\rhom)\times
      \plp}
    {p(\yb_i \mid \deltab, \thb_i)\times p(\thb_i \mid \Rb_i) \times
      p(\rho_i=h \mid \Rb_i, \xibm,\rhom)\times
     \pl} 
    \times \\
  &\hspace{1.3cm}\times\ \frac{
 p(\yb_i \mid \thb_i,\deltab)\times p(\thb_i \mid \Rb_i) \times \gl \times p(\rho_i=h \mid \Rb_i, \xibm,\rhom)
  }{
 p(\yb_i \mid \tbt_i,\deltab)\times p(\tbt_i \mid \Rbt_i) \times \glp \times p(\rhot_i = h'\mid \Rbt_i, \xibm,\rhom)
  }   \Bigg\}=  \\
&=\min\left\{\frac{\plp}{\pl}\times\frac{\gl}{\glp}\right\}.
\end{split}
\end{equation}
Finally, note that to complete the step,
If $\rho_i=H\mm+1$ is accepted, then a new value of $\xibs_{H\mm+1}$
is imputed using the conditional distribution listed below.
\end{itemize}

\subsection*{Update of $\deltab$}

Vector $\deltab$ is updated using a standard data augmentation step for probit static parameters.
recall previus notation with $\gamma_{i,t}=\zb_{i,t}^\top\thb_{i,t}$. Thus introduce:
\[y_{i,t} =
\begin{cases}
1 & \text{if } \zeta_{i,t} > 0 \\
0 & \text{if } \zeta_{i,t} \leq 0
\end{cases}
\]
with $\zeta_i=\xb_{i}^\top\deltab+\gamma_{i,t}+\varepsilon_{i,t}$. Denote with $\mathcal{Z}=\{\zeta_{i,t}: i=1,\ldots,N,\  t=1,\ldots,T_i\}$ the set of all random introduced random variables. We thus derive:
\[p(\mathcal{Z}, \deltab\mid \Theta,\mathcal{D})=\left[\ \prod_{i=1}^N\prod_{t=1}^{T_i}p(\zeta_{i,t}\mid \gamma_{i,t},\deltab)\cdot p(y_{i,t}\mid \zeta_{i,t})\right]p(\deltab) \]
Then we can sample from the full conditional distribution of $\deltab$. 
First for all non-missing value of $y_{i,t}$ perform the data augmentation step as.
\begin{align*}
\text{if } y_{i,t}=1 \text{ sample } \zeta_{i,t}\sim \mathcal{TN}(\xb_i^\top\deltab+\gamma_{i,t},1,\ 0,+\infty) \\ 
\text{if } y_{i,t}=0 \text{ sample } \zeta_{i,t}\sim \mathcal{TN}(\xb_i^\top\deltab+\gamma_{i,t},1,\ -\infty,0) \\ 
\end{align*}
Denote with $\Gamma$ an ordered vector of $\gamma_{i,t}$ for all observed $y_{i,t}$. Similary define $\zb_{all}$, $X_{all}$ the vector for all $\zeta_{i,t}$ and the design matrix for all data that match the correct order used in $\Gamma$. So that we can sample new values of
\[\deltab\mid\Gamma,\zb_{all}\sim\mathcal{N}(\boldsymbol{m}_{post},V_{post})\]
with 
\[V_{post} = (X_{all}^\top X_{all} + D_0^{-1})^{-1}, \quad m_{post} = V_{post}^{-1} (X^\top_{all} (\zb_{all}-\Gamma) + D_0^{-1} d_0)
\]

\subsection*{Update of $\boldsymbol{\xi_{h}^*}$}
The model is conditionally conjugate for $\xibs_h$. We use sampling
from the complete conditional posterior.
Let $n_{h,\ell} =
\sum_{i=1}^n\II(\rho_i=h)\times\II(\Rb_i=r_\ell)$. Then \ech
% Conjugate \bch updating for $\xibs_k$: \ech
% is available for $\xib_i$ using infinite mixture representation of
% $\mbox{CRP}$ (e.g., see \citealp{neal2000markov}).  
\begin{multline}
p\big(\xibs_h\mid\deltab,\{\tb_i\}_{i=1}^n,\{\rho_i\}_{i=1}^n\big)\propto\ 
p\big(\xibs_h,\deltab,\{\tb_i\}_{i=1}^n,\{\rho_i\}_{i=1}^n\big)\\
= \ 
p\big(\xibs_h,\deltab,\{\tb_i\}_{i=1}^n,\{ \Rb_i\}_{i=1}^n,\{\rho_i\}_{i=1}^n\big)\\
=\ 
p\big(\deltab,\{\tb_i\}_{i=1}^n,\{ \Rb_i\}_{i=1}^n,\{\rho_i\}_{i=1}^n\mid \xibs_h\big)p(\xibs_h)\\
\propto\   \prod_{i=1}^n\prod_{\ell=0}^{L-1} \Big[p( \tb_i \mid
           \Rb_i=r_\ell)\  p(\Rb_i=r_\ell\mid\xibs_h)\Big]^{
           \big\{\II(\rho_i=h)\ \times\
           \II(\Rb_i=r_\ell)\big\}}p(\xibs_h) \ p(  \deltab)\\
           \propto 
\prod_{\ell=0}^{L-1} \left(\xis_{h,\ell}\right)^{n_{h,\ell}} p(\xibs_h) 
% \propto\ &  \prod_{i=1}^n \prod_{\ell=1}^L
% \Big[p(\Rb_i=r_\ell\mid\xibs_k)\Big]^{ \big\{\II(\rho_i=k)\ \times\
% \II(\Rb_i=r_\ell)\big\}}p(\xibs_k)   \\ 
% \propto\ &  \prod_{i=1}^n \prod_{\ell=1}^L \big[\xi_{k\ell}\big]^{ \big\{\II(\rho_i=k)\ \times\ \II(\Rb_i=r_\ell)\big\}}p(\xibs_k)   \\
\nonumber
\end{multline}
and therefore 
\begin{equation}\xibs_h \mid \cdot \sim
  \Dir\left( a_\ell+  n_{h,\ell}:  \ \ell = 0,\dots,L-1\right)
%    \sum_{i=1}^n\II(\rho_i=h)\times\II(\Rb_i=r_\ell)\ : 
\end{equation}

\newpage

\section{Priors and Events of Interest}\label{apx:B}
\subsection{Unscaled Model Priors}
Let  $\mathcal{T}^\star_i$ denote the set of time points when treatment
change occurs for subject $i$ and define $\Delta(t)$ as the time (in days) since the last treatment change. \\ 
For $t\not\in\mathcal{T}^\star_i$:
\begin{equation}G_t=G=\left[\begin{matrix}
  1   & 0 & 0   &  \dots &  0 &0\\
    0   & 1 &  0   & \dots &  0 &0\\
 0   & 0   & 0& \dots & 0&0\\
  0   &0  & 1  &  \dots &  0 &0\\
\vdots& \vdots &\vdots   & \ddots  & \vdots  &\vdots \\
 0   & 0 & 0  & \dots  &1 &0 \\
 \end{matrix} \right]
\hspace{.5cm} W_t=W_\varepsilon=\left[\begin{matrix}
\frac{10^{-3}}{3} & 0   &0  & 0 &  \dots &  0\\
0&\frac{10^{-3}}{365^2} & 0   & 0 &  \dots &  0\\
  0&0  &10^{-2}  & 0 &  \dots &  0\\
 0 &0  &0 &10^{-4} &  \dots &  0 \\
 \vdots &\vdots & \vdots &\vdots   & \ddots  & \vdots \\
0&0   & 0 &0 & \dots  &10^{-4} \\
 \end{matrix}\right]\quad \zb_{i,t} =\left[\begin{matrix}
1\\
\Delta (t)\\
1\\
\vdots\\
1
 \end{matrix}\right] 
\end{equation}
For all $t\in\mathcal{T}^\star_i$:
\begin{equation}G_t=G^\star=\left[\begin{matrix}
 1/12  & \dots &  1/12 \\
\vdots &\ddots & \vdots   \\
0 &\dots  & 0 \\
 \end{matrix} \right]
\hspace{.5cm} W_t=W^\star=S_0
\end{equation}
With:
\begin{equation}
m_0=\left[\begin{matrix}
0\\
\vdots\\
0\\
 \end{matrix}\right] ,\hspace{.5cm} S_0=\left[\begin{matrix}
10^{-4}  & 0 & 0 &  \dots &  0\\
 0  & 10^{-4} & 0 &  \dots &  0 \\
0 & 0 & 10^{-2} & \dots  & 0\\
0 & 0 & \vdots & \ddots  & 0\\
 0& 0 & 0  & \dots  &10^{-2} \\
 \end{matrix} \right]
 \end{equation}

\subsection{Definitions of summary statistic }
The summary statistic $\Rb_i$ is determined by three events of clinical interest. Let $r_{1,i}, r_{2,i}, r_{3,i}$
denote indicators for the three events, defined below.
Then $\Rb_i \in \{0,\ldots,L-1\}$ is defined based on the $L=8$
possible combinations of events.

Let $\gamma_{i,t}=\zb_{i,t}^\top\tb_{i,t}$. We define
\[r_{1,i}=\begin{cases}
   \quad 1 \quad \text{ if }&\qquad \dfrac{\sum_{t=1}^{T_i} \gamma_{i,t}}{T_i}\geq 1\\
\quad0\quad \text{ otherwise }  
    \end{cases}\]
and aims to identify subjects whose dynamic random effects
consistently lead to higher risk the
overall mean $\mu_i$.  
The presence of \bchGP $r_{i,1}$ \ech intends to limit the shrinkage effect of individual trajectories toward the mean value.
The second indicator is defined as
P\[r_{2,i}=\begin{cases}
           \quad 1 \quad \text{ if } & \qquad \dfrac{\#\{\gamma_{i,t}:\ \gamma_{i,t} \geq \Phi^{-1}(0.95)\}}{\#\{\gamma_{i,t}:\ \gamma_{i,t} \geq 1\}+1}\geq 0.5\\
\quad0\quad \text{ otherwise }  
    \end{cases}\]
and is the indicator for individuals whose at-risk periods are likely to imply a high-risk level.
Compared with the $r_{1, i}$, \ech the role is similar,
but $r_{2,i}$ focuses on local behavior in contrast with the entire subject's time series.  
$R_{2, i}$ is intended to (potentially) inflate the presence of clumping subjects. 
Finally, $r_{i, 3}$ compares the period before the last treatment change with the entire period under the last active treatment.
This is done through two sample averages defined over $\gamma_{i,t}$. 
The $\bar T_{post}$ statistic is the mean of the dynamic component
under the last treatment, while $\bar T_{pre}$ is the mean over
the pre-treatment period.
The number of days considered by the statistic depends on the subjects, i.e.,  considering the maximum between 90 preceding days (three months) and the number of days number of days before the treatment change.
\[ r_{3,i}=\begin{cases}
    \quad 1 \quad \text{ if }        & \qquad \bar T_{pre}\leq  \bar T_{post}\\
    \quad 0 \quad \text{ otherwise } 
\end{cases}
\]
With:
\[
\begin{split}
&\bar T_{pre}= \sum\limits_{t\in \mathcal{T}_{pre}} \dfrac{\gamma_{i,t}}{\#\mathcal{T}_{pre}}\qquad\text{ with }\mathcal{T}_{pre}=\{t:\ t\in[max(t^\circ_i-90,1),\  t^\circ_i ] \}\\
&\bar T_{post}= \sum\limits_{t\in \mathcal{T}_{post}} \dfrac{\gamma_{i,t}}{\#\mathcal{T}_{post}}\qquad\text{ with }\mathcal{T}_{post}=\{t:\ t\in[t^\circ_i,T_i] \}
\end{split}
\]
\bchGP $r_{3, i}$ \ech aims to identify subjects that likely will require a change of treatment as it captures the evidence for being a treatment non responder subject.

\newpage

\section{Supplementary Analyses and Posterior
  Summaries}\label{apx:C} 
\subsection{Scenarios for the Simulations Study}
Simulations were set up to validate the ability to identify
meaningful clusters among unaligned binary time-series data and to validate posterior inference for the clinically relevant events.

\begin{figure}[H]
    \centering
    \includegraphics[width=\linewidth]{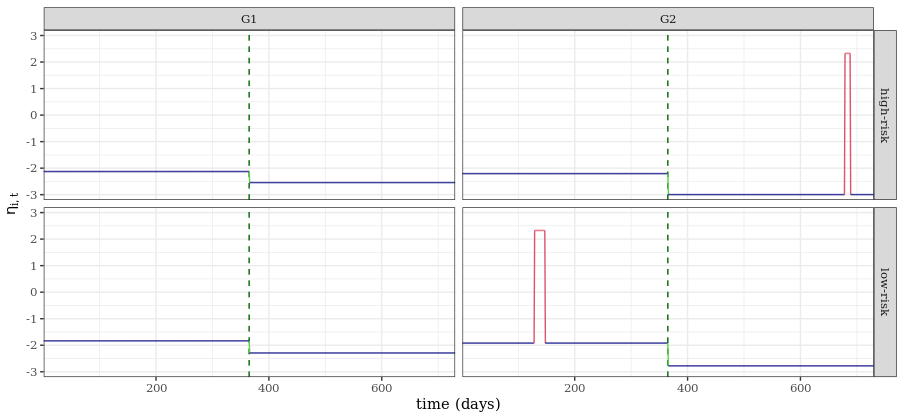}
    \caption{Simulation Truth. The figure represents the latent state
      truth (on a probit scale) used to simulate the binary time
      series. 
    Columns report the latent state for the two clusters of subjects
    in the data.   Rows show the difference between synthetic low- and
    high-risk subjects.  
    Clumping periods are marked in red.
    Length and timing of the clamping periods are random and
    simulated differently for each subject and simulated dataset. } 
    \label{fig:sim_t}
\end{figure}

Synthetic data sets were generated to mimic the heterogeneity and
complexity observed in the data from the Human Epilepsy Project
($\mbox{HEP}$).  Each dataset consisted of a binary time series
representing seizure occurrences, along with associated covariates.
Two distinct clusters of subjects were simulated, each characterized
by different behavioral patterns of seizure occurrence.  Within each
cluster, subgroups were further defined by variations in the static
linear predictors to reflect baseline heterogeneity.  Simulated
trajectories covered a follow-up period of two years, with treatment
changes occurring at the end of the first year.  The binary outcomes
were generated using the latent states shown in Figure \ref{fig:sim_t}.
The corresponding probabilities (of seizures) are tabulated in
% The probabilities of seizures were generated with values reported in
Table \ref{sim:values}. The probabilities imply 
$\Pr(\Rb_i=0)=1$ for the first  group (G1), and
$\Pr(\Rb_i=2)=1$ for the second group (G2).
Periods of clumping were simulated by randomly choosing durations
ranging from 7 to 31 days while within the "clumping" latent state.
Seizure events were then generated based on the probability
$\Pr(y_{i,t}=1) = 99/100$, ensuring a consistent presence of the
events.

Binary time-series data were generated for 100 subjects in each
simulation, with 50 individuals assigned to each cluster. 
Finally, 10\% of data was censored, to mimic missing data.

\begin{table}[H]
    \centering
    \begin{tabular}{l r r r r}
             & \multicolumn{2}{c}{Non-clumping (G1)} & \multicolumn{2}{c}{Clumping (G2)} \\
\midrule
& $1^{st}$ year & $2^{nd}$ year & $1^{st}$ year & $2^{nd}$ year\\
    \midrule
   low-risk  &   1 /60 &  1/365   &   5 /365 & 1/730   \\  
   high-risk &   2 /60 &  4/365   &   10/365 & 1/365   \\[.1cm]
\midrule
\end{tabular}
    \caption{Probabilities used to simulate events in the
      simulation. The values on y-axis in Figure \ref{fig:sim_t}
      correspond to the inverse normal CDF of the values reported in
      this table. } 
    \label{sim:values}
\end{table}

% Figure \ref{fig:bias} reports the absolute bias for the static parameters. The figure shows that the inference model is slightly better in modeling the effects, yet with a significant increase in variance.
% In contrast, evaluating the Dynamic evolutions jointly with other static parameters, make the static intercept estimate significantly worse for under the fully Bayesian approach. This is probably due to a lack of parameters identifiability between the dynamic and static- intercept.
% 
% \begin{figure}
%     \centering
%     \includegraphics[width=0.5\linewidth]{FIGURES/bias_v2.pdf}
%     \caption{Bias over simulations. Figure reports absolute Bias over 100 simulated datasets for the two-step approach ($g(x)$) and for the proposed inference model ($p(x)$). The two-step use maximum likelihood estimates under a static probit model. The inference model evaluates the bias with posterior expected value.}
%     \label{fig:bias}
% \end{figure}

\subsection{Supporting Information on HEP1 Data Analysis}
In this section, we summarize the posterior distribution on
$\deltab$; show an exploratory analysis of treatment effects by
suitable posterior summaries, and marginal fits for all subjects.

This section presents summary statistics of the posterior $\deltab$ from the static parameter inference, exploratory analysis of treatment effects, along with the results of the marginal inference for all the subject parameters derived from the joint inference of the dynamic model.

Table \ref{tab:HEP_delta} summarizes the posterior distribution for
the static regression parameters $\deltab$. Each column of the table summarizes the
marginal posterior of one regression coefficient $\delta_j$. The rows
marked Lower/Upper HDI report a 95\% posterior credible
intervals.

\underline{Pre-defined events of interest:}
Table \ref{tab:long_tab} reports marginal inference for some randomly
chosen subjects.
For each patient, we report the marginal posterior probability of the
pre-defined events of interest.
% Marginal inference was conducted to express the
% probability distributions of clinically relevant patterns for
% individual subjects, thereby enabling a comprehensive evaluation of
% the model's efficacy within the population and yielding insights into
% the disease's dynamic characteristics.
Negligible
probabilities less than $0.01$ are replaced by blanks. The
extended table outlines the posterior probabilities for $\Rb_i=\ell$
for each $\ell=0,\ldots,7$ under (i) the reference model
($\g(\Rb_i=\ell\mid\DD)$ and (ii) the inference model
$p(\Rb_i=\rell\mid\DD)$. In the last line for each patient, the table
lists (iii) the posterior expected values for $\xi_{i,\ell}$, i.e. $\mathbb{E}[\xib_i\mid\DD]$).
% as a descriptive summary, illustrating the effect of other data on
% the sample unit.
%A complete table with results for all patients is  available at\hyperlink{https://drive.google.com/drive/folders/1rdmLaHzgc2ZfLpYtGhy5yJwe1EjjTN8w?usp=drive_link}{link}.

\underline{Treatment effects:}
Table \ref{HEP:TRT} reports posterior summaries constructed to
provide evidence for treatment effects. 
The table are constructed as follows.
We collected $R=2500$ posterior samples of
$\boldsymbol{\gamma}^{(r)}_i=\{ \gamma_{i,t}^{(r)}, i=1,\ldots,T_i\}$
with the superindex $r$ indexing the Monte Carlo samples.
Next, the trajectories of all subjects were
split into intervals corresponding to unique treatments. 
For each slice we
evaluated $\hat T_{pre}$ and $\hat T_{post}$ defined similar to
the definition of
$\bar
T_{pre}$ and $\bar T_{post}$ for the currently assigned treatment in the Online Supplementary Material \ref{apx:B} (for the last active treatment, the
definition is identical).
Additionally, for each slice,
we calculated least squares estimates for slope and intercept in a linear
regression for
\[\gamma^{(r)}_{i,t}=\hat a + \hat b\cdot \Delta(t).\]

Table \ref{HEP:TRT} presents the average of these summary statistics
calculated across all posterior Monte Carlo samples and all
time-slices corresponding to the same treatment. %\cut
% Note that as $\Delta(t)$ starts from
% 0 for all the linear regression, $\hat a$ has a meaningful
% interpretation (i.e., the average expected values among the
% treatment).
The last column gives the number of such summary statistics that are
in the average, and the 3rd and 2nd last column give the proportion of
slices with $\hat T_{pre}<\hat T_{post}$ and $\hat b<0$.
The last three blocks in the table report the summaries restricted to
clusters 1 through 3.

\newpage

\begin{table}[H]
\centering
\resizebox{\textwidth}{!}{
\begin{tabular}{lcccccc}
\toprule
&    \multicolumn{1}{c}{Intercept} &   \multicolumn{1}{c}{Sex (Male)} &  \multicolumn{1}{c}{Age (Year)} &   \multicolumn{1}{c}{Abnormal MRI (Yes)} & \multicolumn{1}{c}{Injury (Yes)} &       \multicolumn{1}{c}{Family history of seizures (Yes)}\\ 
\midrule  
Min.              & -1.812   & -0.400   & -0.005   & 0.176   & -0.237  & -0.088   \\ 
1st Qu.           & -1.666   & -0.262   & -0.001   & 0.296   & -0.127  &  0.065   \\ 
Median            & -1.628   & -0.234   &  0.000   & 0.332   & -0.098  &  0.096   \\ 
Mean              & -1.629   & -0.234   &  0.000   & 0.332   & -0.100  &  0.093   \\ 
3rd Qu.           & -1.591   & -0.206   &  0.001   & 0.367   & -0.072  &  0.122   \\ 
Max.              & -1.458   & -0.094   &  0.005   & 0.493   &  0.018   & 0.239   \\
\midrule
Lower HDI (95\%)  & -1.737   & -0.308   & -0.003   & 0.234   & -0.179  &  0.004   \\
Upper HDI (95\%)  & -1.524   & -0.153   &  0.003   & 0.433   & -0.026  &  0.175   \\
   \bottomrule
\end{tabular}}
\caption{Posterior summaries for $\deltab$. The first part of
  the table are marginal posterior summaries for $\delta_j$
  corresponding to different baseline covariates. The bottom part of
  the table provides marginal posterior credible intervals (highest
  posterior density intervals). }
\label{tab:HEP_delta}
\end{table}

\newpage

\begin{table}[H]
\begin{tabular}{llrrrrrrrr}
\toprule
\multicolumn{1}{c}{ $\mbox{ID}$} & & $\Rb_i = 0$ & $\Rb_i = 1$ & $\Rb_i = 2$ & $\Rb_i = 3$ & $\Rb_i = 4$ & $\Rb_i = 5$ & $\Rb_i = 6$ & $\Rb_i = 7$ \\ 
\midrule
\ALB0001 & $\g(\Rb_i=\ell\mid\DD)$ & 0.10 & 0.90 & \zerop & \zerop & \zerop & \zerop & \zerop & \zerop \\  [.1 cm]
        & $p(\Rb_i=\rell\mid\DD)$ & 0.88 & 0.11 & \zerop & \zerop & \zerop & \zerop & \zerop & \zerop \\ 
        & $\mathbb{E}[\xib_i\mid\DD]$ & 0.86 & 0.06 & 0.02 & 0.01 & \zerop & \zerop & 0.02 & 0.02 \\   [.4 cm]
\ALB0002 & $\g(\Rb_i=\ell\mid\DD)$ & 0.83 & 0.16 & 0.01 & \zerop & \zerop & \zerop & \zerop & \zerop \\  [.1 cm]
        & $p(\Rb_i=\rell\mid\DD)$ & 0.95 & \zerop & 0.05 & \zerop & \zerop & \zerop & \zerop & \zerop \\ 
        & $\mathbb{E}[\xib_i\mid\DD]$ & 0.91 & \zerop & 0.04 & 0.01 & \zerop & \zerop & 0.01 & 0.02 \\   [.4 cm]
\ALB0003 & $\g(\Rb_i=\ell\mid\DD)$ & 0.53 & 0.46 & \zerop & 0.01 & \zerop & \zerop & \zerop & \zerop \\  [.1 cm]
        & $p(\Rb_i=\rell\mid\DD)$ & 0.95 & 0.03 & 0.01 & 0.01 & \zerop & \zerop & \zerop & \zerop \\ 
        & $\mathbb{E}[\xib_i\mid\DD]$ & 0.91 & 0.02 & 0.02 & 0.01 & \zerop & \zerop & 0.02 & 0.02 \\ [.4 cm]  
\vdots\\ [.4 cm]
\ALB0007 & $\g(\Rb_i=\ell\mid\DD)$ & 0.07 & \zerop & 0.18 & \zerop & 0.04 & \zerop & 0.72 & \zerop \\  [.1 cm]
        & $p(\Rb_i=\rell\mid\DD)$ & 0.01 & \zerop & 0.58 & \zerop & \zerop & \zerop & 0.41 & \zerop \\ 
        & $\mathbb{E}[\xib_i\mid\DD]$ & 0.18 & 0.02 & 0.40 & 0.02 & 0.01 & 0.01 & 0.31 & 0.05 \\   [.4 cm]
\vdots\\ [.4 cm]
\NYU0052 & $\g(\Rb_i=\ell\mid\DD)$ & 0.25 & 0.75 & \zerop & \zerop & \zerop & \zerop & \zerop & \zerop \\  [.1 cm]
        & $p(\Rb_i=\rell\mid\DD)$ & 0.94 & 0.06 & \zerop & \zerop & \zerop & \zerop & \zerop & \zerop \\ 
        & $\mathbb{E}[\xib_i\mid\DD]$ & 0.91 & 0.03 & 0.02 & 0.01 & \zerop & \zerop & 0.02 & 0.02 \\   [.4 cm]
\vdots\\ [.4 cm]
\UAB0015 & $\g(\Rb_i=\ell\mid\DD)$ & 0.59 & 0.41 & \zerop & \zerop & \zerop & \zerop & \zerop & \zerop \\  [.1 cm]
        & $p(\Rb_i=\rell\mid\DD)$ & 0.95 & 0.01 & 0.03 & 0.01 & \zerop & \zerop & \zerop & \zerop \\ 
        & $\mathbb{E}[\xib_i\mid\DD]$ & 0.91 & 0.01 & 0.02 & 0.01 & \zerop & \zerop & 0.02 & 0.01 \\   [.4 cm]
\vdots\\ [.4 cm]
\UCSF0027 & $\g(\Rb_i=\ell\mid\DD)$ & 0.02 & \zerop & 0.98 & \zerop & \zerop & \zerop & \zerop & \zerop \\  [.1 cm]
        & $p(\Rb_i=\rell\mid\DD)$ & 0.01 & \zerop & 0.99 & \zerop & \zerop & \zerop & \zerop & \zerop \\ 
        & $\mathbb{E}[\xib_i\mid\DD]$ & 0.20 & 0.02 & 0.62 & 0.02 & 0.01 & 0.01 & 0.07 & 0.06 \\ [.4 cm]
$\vdots$\\[.4 cm]
\bottomrule
\end{tabular}
\caption{
Marginal inference results for example subjects. The table presents posterior probabilities of clinically relevant patterns for individual subjects, aiding in model evaluation and capturing disease dynamics.
Negligible probabilities ($<0.01$) were excluded for clarity. 
Table includes posterior probabilities for events of interest under both the reference model (i.e., $\g(\Rb_i=\ell\mid\DD)$) and the inference model (i.e.,$p(\Rb_i=\rell\mid\DD)$), along with posterior expected values (i.e., $\mathbb{E}[\xib_i\mid\DD]$). 
The full table is available at the provided link.
The first tree subjects correspond to the head of the full table.
Remaining subjects are the 4 example subjects selected in the main manuscript.}
\label{tab:long_tab}
\end{table}

\newpage

% latex table generated in R 4.4.2 by xtable 1.8-4 package
% Sun Mar  9 13:17:20 2025
\begin{table}[H]
\centering
\resizebox{.95\textwidth}{!}{
\begin{tabular}{lccccccr}
\toprule
& \multicolumn{6}{c}{All subjects} & \\[.1cm]
&$\hat T_{pre}$ & $\hat T_{post}$ & intercept & slope & Prop. of $\hat T_{pre} < \hat T_{post}$ & Prop. of $\hat b  <0$ & N. of slices\\  
  \midrule
Carbamazepine & -0.0979 & -0.6908 & -0.3223 & -0.0015 & 0.2195 & 0.3332 & 25 \\ 
Carbamazepine \& Levetiracetam & 0.9653 & 0.5233 & 0.4552 & 0.0071 & 0.3239 & 0.6213 & 12 \\ 
  Lacosamide & 0.3139 & 0.1950 & 0.1000 & 0.0008 & 0.4519 & 0.4748 & 22 \\ 
  Lamotrigine & 0.0715 & -0.6711 & -0.1048 & -0.0017 & 0.1426 & 0.2892 & 129 \\ 
  Lamotrigine \& Lacosamide & 0.8896 & 0.3749 & 0.5139 & 0.0004 & 0.1951 & 0.4352 & 15 \\ 
  Lamotrigine \& Levetiracetam & 0.0428 & 0.1129 & 0.1219 & -0.0020 & 0.4816 & 0.4553 & 64 \\ 
  Lamotrigine \& Oxcarbazepine & 0.7140 & 0.4171 & 0.2874 & 0.0025 & 0.3137 & 0.5993 & 12 \\ 
  Levetiracetam & 0.0171 & -0.4401 & -0.1829 & 0.0002 & 0.2899 & 0.4008 & 171 \\ 
  Levetiracetam \& Lacosamide & 0.6477 & 0.1814 & 0.4426 & 0.0014 & 0.3166 & 0.3983 & 19 \\ 
  Levetiracetam \& Oxcarbazepine & 0.5775 & -0.1738 & 0.0377 & -0.0016 & 0.1993 & 0.3695 & 27 \\ 
  Oxcarbazepine & -0.0007 & -0.5257 & -0.0275 & -0.0010 & 0.2215 & 0.2996 & 55\\ 
  Topiramate & -0.0702 & -0.5232 & -0.2372 & -0.0023 & 0.1548 & 0.3142 & 13 \\ 
  Zonisamide & -0.0370 & -0.5222 & -0.2308 & 0.0005 & 0.2801 & 0.3677 & 18 \\ 
\bottomrule
\\[.2cm]
\toprule
& \multicolumn{6}{c}{Cluster 1} & \\[.1cm]
&$\hat T_{pre}$ & $\hat T_{post}$ & intercept & slope & Prop. of $\hat T_{pre} < \hat T_{post}$ & Prop. of $\hat b  <0$ & N. of slices\\  
\midrule
Carbamazepine & -0.2043 & -1.0664 & -0.4426 & -0.0022 & 0.1252 & 0.2482 & 21 \\ 
  Carbamazepine \& Levetiracetam & 0.1066 & -0.5930 & -0.1626 & -0.0025 & 0.2832 & 0.3111 & 6 \\ 
  Lacosamide & 0.0775 & -0.4641 & -0.2498 & -0.0010 & 0.3166 & 0.3720 & 16 \\ 
  Lamotrigine & -0.0880 & -0.9961 & -0.2916 & -0.0022 & 0.0777 & 0.2184 & 111 \\ 
  Lamotrigine \& Lacosamide & 0.3174 & -0.1033 & 0.0080 & 0.0006 & 0.1987 & 0.4754 & 10 \\ 
  Lamotrigine \& Levetiracetam & -0.1999 & -0.2622 & -0.1468 & -0.0027 & 0.4606 & 0.4069 & 53 \\ 
  Lamotrigine \& Oxcarbazepine & 0.2433 & -0.2368 & -0.0620 & -0.0021 & 0.2679 & 0.4034 & 7 \\ 
  Levetiracetam & -0.0882 & -0.7392 & -0.2680 & -0.0015 & 0.2137 & 0.3176 & 147 \\ 
  Levetiracetam \& Lacosamide & 0.1314 & -0.7187 & -0.2205 & -0.0050 & 0.1083 & 0.2548 & 12 \\ 
  Levetiracetam \& Oxcarbazepine & 0.2222 & -0.3773 & -0.1167 & -0.0018 & 0.1651 & 0.3582 & 22 \\ 
  Oxcarbazepine & -0.2077 & -0.9618 & -0.2745 & -0.0025 & 0.1153 & 0.2049 & 43 \\ 
  Topiramate & -0.1879 & -0.6962 & -0.3683 & -0.0025 & 0.1412 & 0.2692 & 12 \\ 
  Zonisamide & -0.2279 & -0.9191 & -0.5464 & -0.0010 & 0.1835 & 0.3128 & 14 \\ 
   \bottomrule
\\[.2cm]
\toprule
& \multicolumn{6}{c}{Cluster 2} & \\[.1cm]
&$\hat T_{pre}$ & $\hat T_{post}$ & intercept & slope & Prop. of $\hat T_{pre} < \hat T_{post}$ & Prop. of $\hat b  <0$ & N. of slices\\  
\midrule
Carbamazepine & 0.4612 & 1.2811 & 0.3090 & 0.0025 & 0.7144 & 0.7798 & 4 \\ 
  Carbamazepine \& Levetiracetam & 1.8239 & 1.6396 & 1.0730 & 0.0167 & 0.3647 & 0.9316 & 6 \\ 
  Lacosamide & 0.6764 & 2.3792 & 0.9246 & 0.0068 & 0.9819 & 0.9131 & 4 \\ 
  Lamotrigine & 1.3033 & 1.7609 & 1.3550 & 0.0019 & 0.6127 & 0.7372 & 14 \\ 
  Lamotrigine \& Lacosamide & 2.4657 & 1.5987 & 2.0446 & -0.0018 & 0.0549 & 0.1565 & 3 \\ 
  Lamotrigine \& Levetiracetam & 1.0912 & 2.1570 & 1.4249 & 0.0030 & 0.7115 & 0.7544 & 9 \\ 
  Lamotrigine \& Oxcarbazepine & 1.3729 & 1.3325 & 0.7764 & 0.0088 & 0.3779 & 0.8736 & 5 \\ 
  Levetiracetam & 0.6510 & 1.3490 & 0.2679 & 0.0094 & 0.7111 & 0.8980 & 20 \\ 
  Levetiracetam \& Lacosamide & 1.6106 & 1.7074 & 1.5951 & 0.0133 & 0.6214 & 0.5981 & 6 \\ 
  Levetiracetam \& Oxcarbazepine & 2.1257 & 1.5334 & 1.2086 & 0.0026 & 0.5824 & 0.6425 & 3 \\ 
  Oxcarbazepine & 1.0241 & 1.5236 & 1.1861 & 0.0055 & 0.6649 & 0.7160 & 10 \\ 
  Topiramate & 1.3420 & 1.5535 & 1.3363 & 0.0008 & 0.3172 & 0.8544 & 1 \\ 
  Zonisamide & 0.6311 & 0.8670 & 0.8739 & 0.0060 & 0.6182 & 0.5598 & 4 \\ 
\bottomrule
\\[.2cm]
\toprule
& \multicolumn{6}{c}{Cluster 3} & \\[.1cm]
&$\hat T_{pre}$ & $\hat T_{post}$ & intercept & slope & Prop. of $\hat T_{pre} < \hat T_{post}$ & Prop. of $\hat b  <0$ & N. of slices\\  
\midrule
Lacosamide & 1.4801 & 1.0996 & 1.2488 & 0.0028 & 0.4752 & 0.4206 & 2 \\ 
  Lamotrigine & 0.4585 & -0.6620 & 0.1854 & -0.0014 & 0.0392 & 0.4970 & 2 \\ 
  Lamotrigine \& Lacosamide & 1.5991 & 0.3992 & 0.4344 & -0.0004 & 0.0160 & 0.3956 & 1 \\ 
  Lamotrigine \& Levetiracetam & 1.7560 & 0.8532 & 1.3782 & -0.0065 & 0.0016 & 0.3918 & 2 \\ 
  Levetiracetam & 0.6114 & 1.3903 & 0.4365 & 0.0135 & 0.9792 & 0.9653 & 3 \\ 
  Levetiracetam \& Lacosamide & 1.0652 & 1.8265 & 1.4848 & 0.0067 & 0.9868 & 0.9224 & 1 \\ 
  Levetiracetam \& Oxcarbazepine & 2.0353 & -0.4520 & 0.0026 & -0.0073 & 0 & 0.0592 & 1 \\ 
  Oxcarbazepine & -0.5948 & -2.1116 & -0.9333 & -0.0027 & 0 & 0.0780 & 1 \\ 
   \bottomrule\\
\end{tabular}}
\caption{Results of the post-processing of the chain. 
The rows of the table correspond to analyzed subgroups of subjects.
The table reports the average summary statistics computed across MCMC samples and time slices subjected to the same treatment. These include the estimated pre- and post-treatment times, the intercept and slope from standard linear regression, and the proportion of successful treatments based on specific criteria. }
\label{HEP:TRT}
\end{table}
\ech

\end{document}